\pdfoutput=1 %pdflatex?
\documentclass[useAMS,usenatbib]{mn2e}
\usepackage[pdftex]{graphicx} %minipage + width %pdftex for pdflatex?
\usepackage{amssymb}
\usepackage{url}
\providecommand{\apj}{ApJ}

\providecommand{\apjs}{ApJS}
\providecommand{\aap}{A{\&}A}
\providecommand{\aaps}{A{\&}AS}
\providecommand{\araa}{ARA{\&}A}
\providecommand{\aj}{AJ}

\providecommand{\mnras}{MNRAS}
\providecommand{\na}{New Astron.}

\providecommand{\pasp}{Publications of the Astronomical Society of the Pacific}
\providecommand{\rmxaa}{Revista Mexicana de Astronom\'ia y Astrof\'isica}

\newcommand{\Av}{A_{\rm v}}
\newcommand{\avgAv}{\langle A_{\rm v} \rangle}
\newcommand{\HI}{\mbox{H{\sc i}}}
\newcommand{\Htwo}{\mbox{H$_2$}}
\newcommand{\kms}{{\rm km\ s}^{-1}}
\newcommand{\pcc}{{\rm cm}^{-3}}

\title[Synthetic observations \HI\ and molecular gas]{Molecular cloud
formation as seen in synthetic \HI\ and molecular gas observations}
\author[Heiner, V\'azquez-Semadeni \& Ballesteros-Paredes]{Jonathan S. Heiner$^{1}$\thanks{E-mail: jonathanheiner@gmail.com}, Enrique V\'azquez-Semadeni$^{1}$, Javier Ballesteros-Paredes$^{1}$\\
  $^1$Centro de Radioastronom\'ia y Astrof\'isica (CRyA), Universidad Nacional Aut\'onoma de M\'exico, C.P. 58190 Morelia, Michoac\'an, Mexico\\
  }

\begin{document}

\date{Draft of \today}
\pagerange{\pageref{firstpage}--\pageref{lastpage}} \pubyear{201X}
\maketitle
\label{firstpage}

\begin{abstract}
We present synthetic \HI\ and CO observations of a numerical simulation of
decaying turbulence in the thermally bistable neutral medium. We first
present the simulation, which produces a clumpy medium, with clouds
initially consisting of clustered clumps. Self-gravity causes these
clump clusters to merge and form more homogeneous dense clouds. We apply
a simple radiative transfer algorithm, throwing rays in many directions
from each cell, and defining every cell with $\avgAv\ > 1$ as molecular. We
then produce maps of \HI, CO-free molecular gas, and CO, and investigate
the following aspects: i) The spatial distribution of the warm, cold,
and molecular gas, finding the well-known layered structure, with
molecular gas being surrounded by cold \HI\ and this in turn being
surrounded by warm \HI. ii) The velocity of the various components,
finding that the atomic gas is generally flowing towards the molecular
gas, and that this motion is reflected in the frequently observed
bimodal shape of the \HI\ profiles. This conclusion is, however,
tentative, because we do not include feedback that may produce \HI\ gas
receding from molecular regions. iii) The production of \HI\ 
self-absorption (HISA) profiles, and the correlation of HISA with
molecular gas. In particular, we test the suggestion of using the second
derivative of the brightness temperature \HI\ profile to trace HISA and
molecular gas, finding significant limitations. On a scale of several
parsecs, some agreement is obtained between this technique and actual
HISA, as well as a correlation between HISA and the molecular gas column
density. This correlation, however, quickly deteriorates towards 
sub-parsec scales. iv) The column density PDFs of the actual
\HI\ gas and those recovered from the \HI\ line profiles, finding that
the latter have a cutoff at column densities where the gas becomes
optically thick, thus missing the contribution from the HISA-producing
gas. We also find that the power-law tail typical of gravitational
contraction is only observed in the molecular gas, and that, before the
power-law tail develops in the total gas density PDF, no CO is yet
present, reinforcing the notion that gravitational contraction is needed
to produce this component.

\end{abstract}

%mn2e
\begin{keywords}
  turbulence -- ISM: clouds -- ISM: evolution -- ISM: kinematic and dynamics -- ISM: structure
\end{keywords}

\section{Introduction}

%Figure of Sec. 2.1
\begin{figure*}
  \resizebox{0.9\columnwidth}{!}{\includegraphics{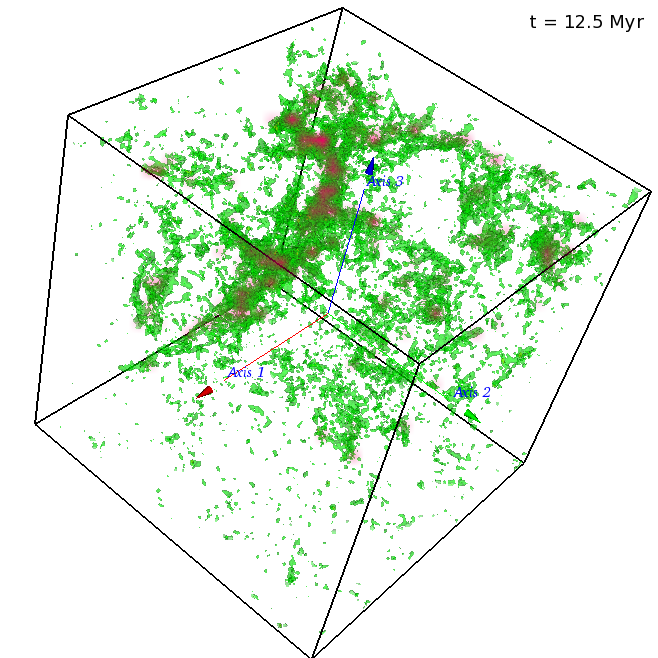}}
  \resizebox{0.9\columnwidth}{!}{\includegraphics{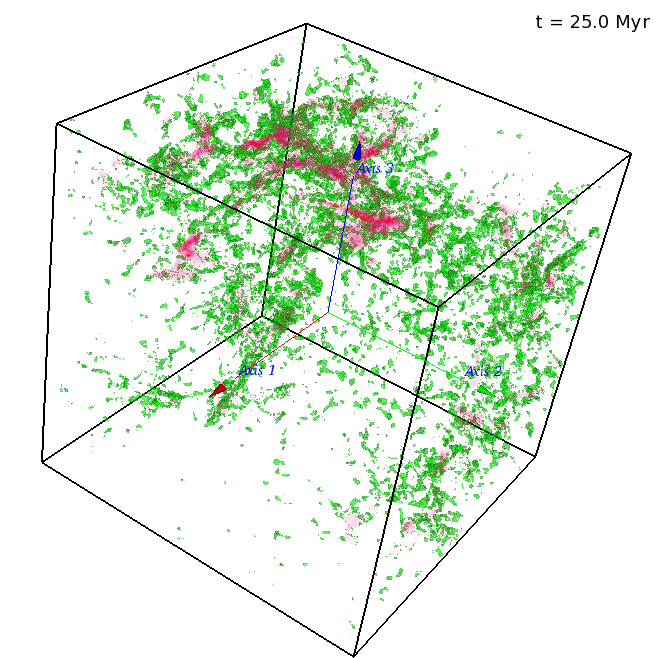}}
  \resizebox{0.9\columnwidth}{!}{\includegraphics{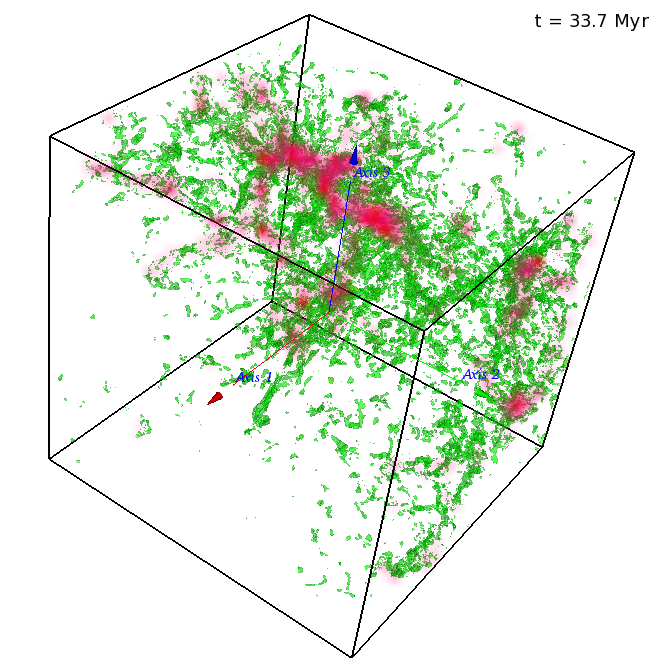}}
  \caption{\label{fig:3dplots} Three-dimensional distribution of the
  atomic density (green) and the molecular density (red) at the three
  timesteps through iso-surface renderings made with Starlink's {\sc GAIA}
  \citep[see e.g.][]{2007ASPC..376..695D}. Two green intensity levels
  were used, of 1 and 3, 1.5 and 3, and 1.5 and 3 $M_{\odot}\
  \rm{pc}^{-3}$, for the timesteps at 12.5, 25.0, and 33.7 Myr,
  respectively. Similarly, red levels of 0.44 and 13, 0.44 and 50, and
  0.44 and 50 $M_{\odot}\ \rm{pc}^{-3}$, were respectively used at each
  of the timesteps to image the molecular gas density. The numbered axes
  correspond to the x, y and z axes respectively (with the x-axis
  pointing towards the lower left corner. \textit{A color
  version this figure is available in the online version of this
  journal.}}
\end{figure*}

The present view of the interstellar medium (ISM) is that it is in
general highly turbulent 
\citep[e.g.,][]{2000ApJ...540..271V,2004RvMP...76..125M,2004ARA&A..42..211E,2007RMxAA..43..123B}, and that dense, cold clouds form where
turbulent compressions or larger-scale instabilities produce converging
flows that in turn cause the density to increase locally \citep[e.g.,][]{2014MNRAS.437L..31D,2014arXiv1402.6196M}. Indeed, numerical simulations of converging
flows in the warm neutral medium (WNM) including self-gravity but no
stellar feedback \citep{2007ApJ...657..870V,2011MNRAS.414.2511V,2008A&A...486L..43H,2008ApJ...689..290H,2009ApJ...704.1735H,2009pjc..book..421B} show in general that, once a dense cloud is formed by this
mechanism, it quickly becomes gravitationally unstable, and begins to
undergo gravitational collapse. An important feature of this collapse is
that it begins in gas that should be primarily atomic, with molecule
formation occurring as a consequence of the gravitational contraction, as
initially proposed on theoretical grounds by \citet{1986PASP...98.1076F} and
\citet{2001ApJ...562..852H}. 
Simulations including a self-consistent treatment of the chemistry
\citep[e.g.,][]{2012MNRAS.424.2599C,2013arXiv1306.5714C} indeed show that this is so, and that
H$_2$ molecule formation occurs relatively early during the collapse,
while CO formation only occurs some 2 Myr before star formation (SF)
starts. Finally, simulations including stellar feedback \citep{2010ApJ...715.1302V,2013MNRAS.tmp.2055C} show that the infalling motions of the dense gas are not
overturned by the action of the feedback. Rather, the clouds are
progressively evaporated, with the escaping gas being warm and diffuse,
while the dense gas continues to fall in.

The above picture of molecular clouds formed by flows implies that CO-identified molecular clouds (MCs) should be surrounded by cold atomic gas, perhaps mixed with CO-dark molecular gas, and that this medium should in turn be embedded in warm
\HI, similarly to the classical picture of the ISM \citep[see, e.g., the
review by ][and references therein]{1993prpl.conf..125B}, except for the additional
property that the atomic and CO-dark components are expected to be
flowing towards the MCs.

Observations partially support
this view, since giant molecular clouds (GMCs), whose largest dimension
reaches up to $\sim 100$ pc, appear to be the densest regions in the
ISM, and are known to be embedded in CO-free molecular gas, which in
turn is embedded in \HI\ superclouds, of sizes of up to 1 or 2 kpc
\citep[see, e.g.,][and references therein]{2014arXiv1402.6196M}. 
In order to distinguish whether molecular clouds are formed by \HI\ flows,
it is necessary to establish the motions of the \HI.
However, observationally establishing the direction of the motions that produce
GMCs is a formidable problem, due to the confusion caused by the
ubiquitous presence of \HI\ gas in the Galactic disk.

%The review by \citet{2004ARA&A..42..211E} reminds us that the idea of a
%turbulent interstellar medium (rather than one with relatively low gas
%velocities) was proposed early on but has only fairly recently gained
%more popularity. Perhaps the advent of high resolution numerical
%simulations resulted in this renewed interest. Turbulent motions at a
%particular scale can provide support for larger clouds, while promoting
%collapse on smaller ones \citep{2006MNRAS.372..443B}. Where the
%molecules form, however, is an entirely different question. For example,
%\citet{2001ApJ...562..852H} claim that molecular gas accumulates in the
%atomic phase, allowing rapid star formation once a high density cloud is
%formed. As noted by \citet{1999ApJ...527..285B}, it appears that
%molecular clouds have a relatively short lifetime based on the spatial
%and age distribution of stars formed inside of it. Therefore, the
%molecular gas needed to form stars has to accumulate rapidly. On the
%other hand, molecular hydrogen may not be needed for star formation at
%all \citep{2012MNRAS.421....9G} in which case any molecular formation
%timescale argument would not apply. At any rate, N-body simulations
%produce realistic realizations of the interstellar medium and as such
%can help us interpret actual observations of the same. It has become
%possible in the last decade to conduct numerical simulations routinely
%and we present the outcome of one such simulation here.

Several studies have been conducted to determine the signatures in line
profiles and position-velocity (PV) space arising from the density and
velocity features produced in the simulations. Early studies simply
investigated line-of-sight (LOS) projections of the density
field from numerical simulations, and perhaps investigated the column
density in the velocity coordinate (line profiles and channel maps)
\citep[e.g.,][] {1999ApJ...527..285B,2000ApJ...532..353P,2001ApJ...546..980O,2002ApJ...570..734B},
although without performing synthetic observations based on
integration of the radiative transfer (RT) equation for the various
lines involved. 
A further step was made by \citet{2002ApJ...570..734B} who,
in order to study the internal structure of molecular clouds, created 
synthetic CO and CS line profiles in local thermal equilibrium from
numerical simulations of isothermal molecular clouds. They found that the
density-size relation, rather than an intrinsic property of molecular clouds,
is an artifact of the observational procedure \citep[see also][]{2012MNRAS.427.2562B}.
More recently, synthetic observations have been
performed to varying degrees of approximation, and used to study to what
extent the actual density and velocity structure of the atomic gas can
be inferred from the line profiles \citep{2007A&A...465..445H}, or to
show that infalling motions in the molecular gas produce realistic CO
spectra: For example by matching the linewidths as well as the magnitude of the velocity dispersions seen in the $^{13}$CO filamentary structure of Galactic molecular clouds \citep{2009ApJ...704.1735H}. 

In this contribution we take one step further in this direction by
combining synthetic observations of atomic {\it and} molecular gas from
a numerical simulation of the formation of dense clouds in the turbulent
ISM. These can help interpreting actual observations of MCs and their
atomic envelopes, and help understanding the atomic-to-molecular transition
in the ISM. Specifically, we use a simple radiative transfer (RT)
algorithm on the output of a numerical simulation of decaying,
self-gravitating turbulence in the ISM, in order to classify the gas as
either being atomic or molecular, and then investigate various aspects,
such as the spatial distribution of the molecular and cold and warm
atomic components, as well as their velocities, and the signatures of
these motions on the line profiles and intensity maps. 

An important issue to assess is the production of \HI\ self-absorption
(HISA) features by the cold atomic gas expected to surround the CO
clouds.  Although HISA is perhaps the most reliable method to detect
cold \HI, it is not free from uncertainties and ambiguities. In
particular, it is important to be able to distinguish between a true
HISA feature, and lack of background emission. 
%focus on reproducing realistic atomic hydrogen
%line profiles from simulations and what these profiles can tell us about
%the underlying gas cloud structures. 
%apparent
%(self-)absorption features in the observed line profiles of atomic
%hydrogen: These have been known since the beginning of 21-cm
%observations, because these features are easily and commonly seen. Since
%Galactic HI emission can be seen in all directions and at a wide range
%of heliocentric velocities, deriving accurate physical quantities from
%these absorption features remains challenging. However, a number of
%general criteria have been established to discern at least with some
%level of confidence whether a depression in a line profile is caused by
%a lack of gas or by HI self-absorption (HISA). 
To this end, \citet{1974AJ.....79..527K} proposed four criteria, namely
i) a fairly narrow dip (less than about 7 km s$^{-1}$) appearing in the
\HI\ spectrum; ii) an \HI\ velocity feature corresponding to a molecular emission
line; iii) a dip appearing `on-cloud' but not on the `off-cloud'
calibration profile, and iv) a slope of the dip steeper than the slope
of the background emission profile to exclude the possibility of a line
profile composed of a double gaussian peak. 
%Based on these criteria,
%individual features were classified in groups reflecting how convincing
%the features were. \citet{1974AJ.....79..527K} generally fitted a single
%gaussian to the overall brightness temperature profile, to the extent
%possible.
%
%%Knapp 1974 also explains why cold atomic gas can be expected in molecular clouds: a) small amounts of HI are left over from molecule formation or b) the HI to H2 abundance is in equilibrium with a dissociating field such as one produced by cosmic rays.
%
Based on later findings, the HISA detection criteria changed somewhat,
most notably since molecular line emission is not always detected when a
molecular cloud is positively identified, for example through optical
extinction. \citet{2000ApJ...540..851G} generalized the steepness
requirement by stating that the profile wings need to be steeper than
what superposition of neighboring emission lines would cause. 
%In other words,
%if the profile could have been caused by two emission lines instead of
%one with HISA, the profile does not qualify. 
Additionally, a certain
amount of small-scale angular structure was required, as well as a certain
minimum background level. 
%They also noted that these criteria introduce
%some selection biases, such as a favorable viewing geometry, enough
%self-absorption to distinguish from emission gaps and a relatively high
%HI column density. Also, because of the narrow linewidth requirement,
%only quiescent HISA can be identified, although if the HISA originates
%from cold gas inside molecular clouds, this is less of an issue. In our
%simulation, we have a weak background caused by low density hot gas in
%our simulation volume, but since we are not limited by observational
%constraints we do not need to require a strong background to identify
%HISA.

%A catalog of HISA detections is provided by \citet{2005ApJ...626..195G}, 
An automated HISA detection algorithm taking into account spatial and
spectral features was developed by \citet{2005ApJ...626..214G}, while
another example was presented by
\citet{2005ApJ...626..887K}, who used both the first and second
derivative (thresholded at certain levels) of the \HI\ profile to find
HISA. They found that 60\% of the HISA they detected coincides with
molecular (CO) gas and proposed that HISA is related to an
atomic-to-molecular phase transition. \citet{2005ApJ...622..938G}
determined that the cold \HI\ gas coincides with $^{13}$CO in five dark
clouds they considered, and their models show that the cold \HI\ in
those structures has densities between 2 and 6 cm$^{-3}$ as compared to
H$_2$ central densities of 800 to 3000 cm$^{-3}$.

Where the absorption coincides with molecular lines and has
approximately the same linewidth, it has been referred to as `\HI\ narrow
self-absorption', or HINSA \citep{2003ApJ...585..823L}. Since strictly
speaking we do not require molecular (CO) emission in order to identify
\HI\ self-absorption, we will refer to these features as HISA regardless
of whether they may be considered `narrow', as a matter of
convenience. 
%However, velocity-wise our HISA profiles do in fact show
%narrow absorption features.

In general, \HI\ profiles typically look bi-modal on or near molecular
clouds and it can be hard to distinguish without additional information
what physical properties are behind this profile shape. 
%For example,
%these profiles could be caused by colliding gas flows. 
In this contribution we therefore also aim to explore the extent to
which we can distinguish true HISA from separate emission peaks, and how
well the HISA is correlated with the presence of molecular
gas at various evolutionary stages of the clouds. 
%Therefore, we produced realistic HI line profiles from simulations
%of turbulent flows, implemented our own algorithm to detect HISA and
%subsequently compare the HISA features to the molecular cloud structures
%present in our simulation.

This paper is structured as follows: First, we introduce the simulation
and the synthetic observations derived from it in Sec.\
\ref{sec:method}, and discuss the density and velocity structure of
the various gas components in real space, emphasizing the general trend
of inflow onto the dense gas in Sec.\ \ref{sec:gral_morph}. Then, in Sec.\ \ref{sec:synth_prof} we
discuss the nature of the synthetic line profiles, and in Sec.\
\ref{sec:HISA_molec} we address the identification of HISA features and
compare these features to those of the molecular gas. Next, we discuss
the structure of the probability density functions of the simulation
and the structure of the gas in the context of colliding gas flows. In
Sec.\ \ref{sec:concl} we
close with a brief discussion and summary of our results.

\section{The method} \label{sec:method}

%Figure of Section 3
\begin{figure*}
  \resizebox{2\columnwidth}{!}{\includegraphics{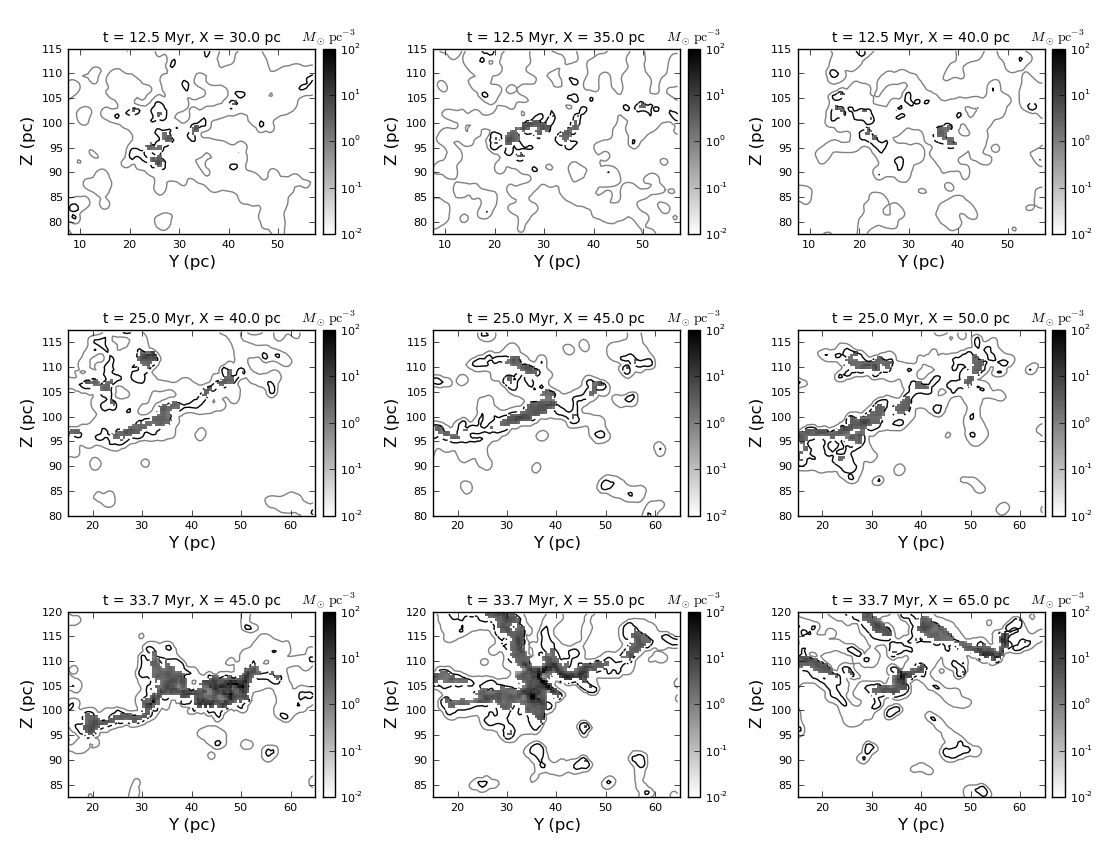}}
  \caption{\label{fig:atom_mol} Slices of the atomic and molecular mass
  density (contours and grayscale respectively) in units of the
  simulation are displayed. The coordinates are in parsec offset with
  respect to one corner of the simulation cube. The atomic density
  contour levels are 0.1, 10 and 100 $M_\odot\ \rm{pc}^{-3}$ (the native
  units of the simulation) with the lowest contour colored gray.}
\end{figure*}

\subsection{The numerical simulation}

Our numerical simulation was performed using the {\sc Gadget-2} code
\citep{2001NewA....6...79S}, with $296^3 \approx 2.6 \times 10^7$ SPH
particles, and including prescriptions for sink particles taken from
 \citet{2005A&A...435..611J} and for heating and cooling from \citet{2007ApJ...657..870V}. The
initial density and temperature of the simulation were set at $3~\pcc$
and $730$ K, respectively, representing the mean ISM conditions at a
spiral arm. The physical size of the numerical domain was 256 pc. The
simulation was started by applying a Fourier turbulence driver with purely
solenoidal modes between wavenumbers $k=1$ and $k=4$ for 0.65 Myr,
reaching a maximum velocity dispersion of $\sigma \approx 18\, \kms$ at
$t \approx 0.65$ Myr.

For the analyses presented in this paper, we consider three timesteps of
the simulation, at 12.5, 25.0 and 33.7 Myr, where the latter is the
final recorded timestep of this simulation. An octant of the simulation,
of size 128 pc per side, containing the most massive clouds, was
interpolated onto a $256^3$ uniform grid, giving an effective resolution
of 0.5 pc. This resolution was chosen to provide a resolution comparable
to observations of the interstellar medium in our Galaxy, while allowing
easy manipulation of the data. Within this 128 parsec$^3$ sub-volume, we
focus on the entire sub-volume as well as on a further zoomed-in volume
containing a dominant cloud feature (not necessarily the same cloud at
each timestep). Images of the sub-volume at the three timesteps, after
applying the RT prescription (cf.\ Sec.\ \ref{sec:synth_obs}) are shown
in Fig.\ \ref{fig:3dplots}.

\subsection{Producing molecular gas} \label{sec:synth_obs}

Once the data has been gridded, we proceed to distinguish between
`atomic' and `molecular' gas.  Note that the version of {\sc Gadget-2}
that we use does not solve the chemistry explicitly, and so we use the
simple criterion proposed by \citet{2008ApJ...689..290H} for deciding
which grid cell in the simulation are `atomic' and which are
`molecular'. While obviously faster than solving a full chemical network
\citep[e.g.,][]{2000ApJ...532..980K, 2012MNRAS.424.2599C}, this prescription is rather
approximative, in particular because it assumes that molecules form as
soon as the physical conditions allow for it, rather than following the
time-dependent process of molecule formation. Thus, our results should
be considered as providing upper limits for the amount of molecular gas
present in the simulation. 
%This should be sufficient for our purposes,
%since we are mostly interested in the presence of molecular gas, not the
%details of its formation.

The prescription we use is as follows: from each grid cell we emit 100
rays in random directions, that reach out to the edge of the numerical
box, and compute the column density along each ray. The column density
is converted to a visual extinction $\Av$ using the standard formula
$\Av \approx 10^{-21} N_{\rm H}$. We then compute the average extinction
$\avgAv$ for each cell. If $\avgAv > 1$ and $T < 50$K for a cell, we label its
contents as molecular, meaning hydrogen is in the form of $\Htwo$
molecules. If, in addition to satisfying the condition $\avgAv > 1$, the
density in the cell satisfies the local condition 
$n_{\rm{H_2}} > n_{\rm crit}$, 
then we assume that the cell contains CO molecules, with a fixed relative
abundance of CO(1-0) to total molecular gas number density of $\sim 2
\times 10^{-4}$ \citep[e.g.][]{1984A&AS...58..327C,2008ApJ...679..481P}.
The critical density is given by
%$n_{\rm crit} \simeq \frac{A_{ji}}{\sigma_{\rm{CO}} \times 9.09 \times 10^3 \times \sqrt{T}}$, 
%
\begin{equation}
n_{\rm crit} \simeq \frac{A_{ji}}{\sigma_{\rm{CO}} \times \left <
v_{\rm{H_2}} \right>},
\end{equation} 
where $A_{ji}$ is the Einstein A, $\sigma_{CO}$ is the collisional cross section 
of the CO(1-0) molecule (ignoring higher transitions), and $\left< v_{\rm{H_2}} \right> \approx
9.09 \times 10^3\times \sqrt(T)$.
The local gas temperature is available from the simulations at each resolution element.
See Sec.\ \ref{sec:synth_prof} for further discussion.

%soon as a resolution element finds
%itself surrounded by an average extinction of $A_v = 1$ or more, and its
%temperature has dropped below 50 K, it is considered molecular. In order
%to determine the mean extinction of a given resolution element, we
%simply calculated the column density in 100 random directions starting
%at that resolution element and computed the average value.

An immediate implication of this procedure is that in our simulation,
atomic and molecular gas are not mixed in a grid cell, which is either
fully atomic, or fully molecular. Nevertheless, we can have mixing of
atomic and molecular gas at scales larger than the grid cell size, if
neighboring cells are in some cases atomic, and in other cases
molecular.  Therefore, it is perfectly feasible to have mixing of atomic
and molecular gas along an observational line of sight although, as
mentioned above, the molecular fraction we estimate will be an upper
limit. In particular, we aim to investigate atomic gas that
is about to turn molecular, and we expect our results, as they relate to \HI\
self-absorption and atomic-to-molecular conversion, to provide us with
an informative, albeit only approximate, view.
%This
%means that we assume that the transition layer between atomic and
%molecular is sufficiently thin (much less than our 0.5 parsec
%resolution) and no significant fraction of cold atomic gas mixed in with
%molecular gas is present. On the other hand, 
Note also that, since we distinguish between atomic and molecular gas by
post-processing of the simulation data, this distinction has no effect
on the evolution of the simulation. We plot the
three-dimensional distribution of the atomic and the molecular gas
qualitatively in Figure \ref{fig:3dplots}.

%After reaching a certain density threshold \textbf{(what mass?)}, gas is
%converted into a point source (sink) particle that can accrete more gas,
%simulating star formation. However, no radiation is modeled from this
%particle nor 
Finally, we note that, because our simulation does not include stellar
feedback, the conversion of atomic gas into molecular gas and finally
into stars is essentially a one-way process, although potentially
molecular gas may become atomic again if the local $A_v$ drops below
unity or if the temperature rises sufficiently. We have not traced
individual particles, so whether, or to what extent, this occurred in
our simulation is unknown. No photodissociation physics were included.

%While having either atomic or molecular gas in a particular volume is,
%ultimately, clearly unrealistic, since even in the densest cloud cores
%it is still expected that some photodissociation takes place due to
%cosmic rays \citep[e.g.][]{1980A&A....91...68D}, this does not
%necessarily mean that the synthetic observational profiles we produce
%are unrealistic. Also, it may seem counterproductive to investigate
%HISA, supposedly caused by cold gas, and then not have the ability to
%form cold gas inside volumes of molecular gas. However, depending on the
%orientation of the cloud, we can still have line-of-sight mixing of
%atomic and molecular gas. Mainly, we can expect to be sensitive to
%atomic gas that is about to turn molecular and our results as they
%relate to HI self-absorption will tell us a partial but informative
%story.

\section{General morphology and cold gas fraction}
\label{sec:gral_morph}

As time progresses in the simulation, filaments and clumps of gas begin
to form by gravitational contraction onto the density peaks produced by
the initial turbulence, aided by the thermal instability
\citep{1999ApJ...527..285B, 2007ApJ...657..870V, 
2009ApJ...707.1023V, 2008ApJ...689..290H, 2009ApJ...704.1735H,
2009MNRAS.398.1082B, 2013arXiv1308.6298G}.
%, which start to dominate the simulation volume.
%EVS_rev1: Strictly speaking, this statement is incorrect, since those
%structures have large densities, and therefore they typically have small
%volumes. 
Figure
\ref{fig:atom_mol} shows slices through the mass density cubes of the
atomic and molecular gas, zoomed in to show example molecular cloud
structures. Molecular clouds are formed in filament-like structures
before collapsing. The positions in parsecs refer to the offset from the
(0,0,0) coordinate at one corner of the gridded cubes.

\begin{figure*}
  \resizebox{2\columnwidth}{!}{\includegraphics{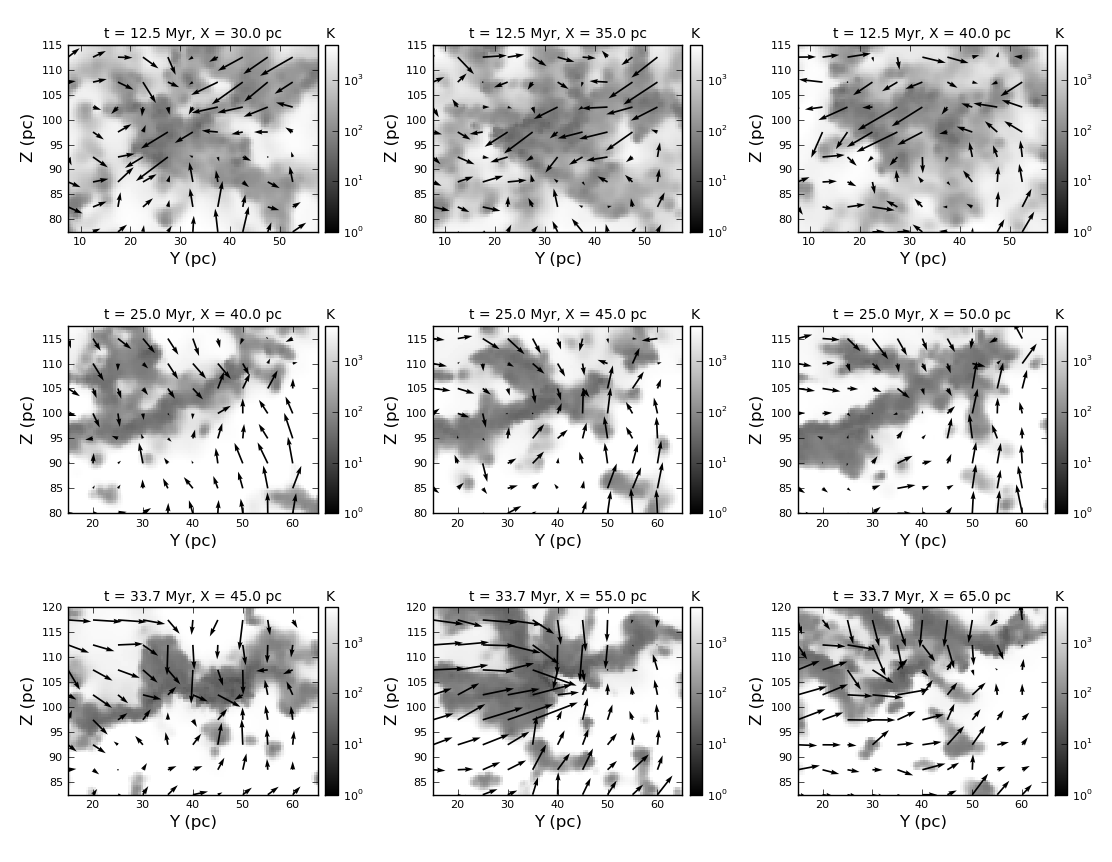}}
  \caption{\label{fig:v_temp} Velocity of the gas in the y, z plane
  (arrows) and the temperature of the gas is plotted. We assume that
  cold gas (below 50K) with a local visual extinction above 1 has turned
  molecular. It can be seen that the gas is flowing into the cold
  areas.}
\end{figure*}

%Figure of Section 4.2
\begin{figure*}
  \resizebox{1.8\columnwidth}{!}{\includegraphics{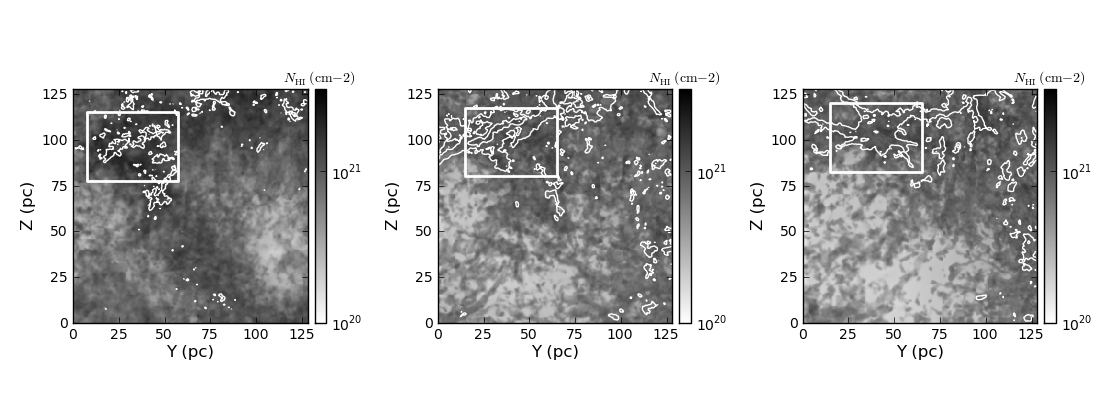}}
  \caption{\label{fig:NHI_Nmol_select} Atomic hydrogen column density
  maps at the three timesteps. The rectangles mark the regions featured in
  subsequent figures (when not considering the full area). The white
  contours represent molecular gas column densities of 3, 5, and 10
  $\times 10^{20}\ \rm{cm}^{-2}$, respectively, in each of the
  timesteps.}
\end{figure*}

In Figure \ref{fig:v_temp}, for the same zoomed-in areas in the three
timesteps, we show the velocity field in the $(y,z)$ direction and the gas
temperature. As expected, comparing this figure with Figure
\ref{fig:atom_mol}, the low density gas has a high temperature (the
equivalent of the warm neutral medium), while denser gas is cooler, and
eventually turns molecular. However, it is also noticeable that there
are {\it cold} \HI\ envelopes around the molecular gas.

%Values calculated with script mass_sigv.py
\begin{table*}
  \centering
  \begin{tabular}{lllllll}
    \hline\hline
    Time       & $M_{\rm{tot}}$ & $M_{\rm{atom}}$ & $M_{\rm{mol}}$ & $\sigma_{v,\rm{tot}}$ & $\sigma_{v,\rm{atom}}$ & $\sigma_{v,\rm{mol}}$ \\
    (Myr)      & ($M_{\odot}$)  & ($M_{\odot}$)   & ($M_{\odot}$)  & (km s$^{-1}$)         & (km s$^{-1}$)          & (km s$^{-1}$) \\
    \hline
    12.5 (full)& $2.0 \times 10^5$ & $1.9 \times 10^5$   & $9.2 \times 10^3$ & 4.5 & 4.5 & 0.23 \\
  (sub-region) & $4.2 \times 10^4$ & $3.8 \times 10^4$   & $3.6 \times 10^3$ & 5.9 & 5.9 & 0.49 \\
\\
    25.0 (full)& $2.3 \times 10^5$ & $1.8 \times 10^5$   & $5.0 \times 10^4$ & 2.7 & 2.7 & 0.20 \\
  (sub-region) & $5.8 \times 10^4$ & $3.8 \times 10^4$   & $2.0 \times 10^4$ & 3.3 & 3.3 & 0.41 \\
\\
    33.7 (full)& $2.7 \times 10^5$ & $1.7 \times 10^5$   & $1.1 \times 10^5$ & 2.2 & 2.2 & 0.28 \\
  (sub-region) & $8.0 \times 10^4$ & $3.2 \times 10^4$   & $4.7 \times 10^4$ & 2.5 & 2.5 & 0.55 \\
    \hline
  \end{tabular}
  \caption{\label{tab:mass_sigv} Masses and velocity dispersions in both
  the gridded volume and the zoomed-in regions.}
\end{table*} 

In Table \ref{tab:mass_sigv} we give the total, atomic, and molecular
masses in the gridded volume as well as in the zoomed-in areas for the
three times we study. It can be seen that the molecular mass fraction
increases rapidly and that the velocity dispersion in the simulation
decreases over time. The average density of the gridded volume increases
slowly, namely 3.0, 3.5 and 4.1 cm$^{-3}$ at each timestep respectively,
where 3.0 cm$^{-3}$ is also the average density of the full simulation
volume. The increasing gas mass and density indicate that gas is flowing
into the gridded volume, which was selected for its developing molecular
cloud structures. This fact is also suggestive of gravitational
contraction at the largest scales we are considering.

%While we have already mentioned that the atomic and molecular gas in our
%simulation do not mix, and that therefore we do not produce cold atomic
%gas inside molecular cores, 
It is also instructive to compute the fraction of cold atomic gas in our
simulation. This cold atomic gas exists when the local $A_v$ has not
reached unity, at which point it would be deemed
molecular. \citet{2010ApJ...724.1402K} found in their observational
survey of molecular cores an abundance of cold \HI\ column density to
total proton column density of $10^{-2.8}$ or $\approx 0.0016$. In our
simulation, we have computed the mass of cold \HI\ (being cooler than 50 K
and having $A_v$ below unity) versus the total gas mass, finding ratios
of 0.00014, 0.00055 and 0.00027 respectively for our three increasing
timesteps, or consistently between half and one order of magnitude less
than the values \citet{2010ApJ...724.1402K} found. This is probably a
consequence of our `instantaneous molecule formation' assumption
which, as mentioned above, causes our molecular gas fractions to be
upper limits.
%Meaning, how much cold gas do we miss relative to Krco and Goldsmith??
Also, this ratio can vary due to molecular clouds forming in a
non-continuous fashion and due to gas entering and leaving the
sub-volume of the simulation that we selected.

\section{Synthetic \HI\ and CO profiles} \label{sec:synth_prof}

\subsection{Generation of the `observations'} \label{sec:profile_generation}

One of the main goals of this contribution is to focus on the
`observer's perspective'. To this end, we derive \HI\ and CO line
profiles. In order to perform the 21-cm \HI\ and CO(1-0) line profiles, we
have assumed that both lines are in local thermodynamic equilibrium
(LTE) for the population of the atom/molecule energy levels. This is a
good approximation for the study of \HI\ profiles, since \HI\ clouds are
dominated by collisions. In the case of molecular clouds, LTE is a
reasonably good approximation to study the CO lines qualitatively, if
the volume density is above a certain excitation threshold \citep{rw96}. 
%We adopted a fixed CO abundance (see below).  

Since we know the detailed
velocity, density and temperature fields, we have integrated the
transfer equation along the line of sight, assuming that every pixel
above the threshold has a blackbody emission at the corresponding
temperature. The details can be found in
\citet{2002ApJ...570..734B}. Here we just note that the \HI\ emission will
come from every resolution element that is not deemed to be molecular,
while the CO emission will come from molecular resolution elements ($A_v
\geq 1$ and $T < 50K$) as well as having a number density above the local
critical density, in order to avoid including
molecules that are under-excited \citep{rw96}. 
%This critical density
%is generally several hundred $H_2$ molecules cm$^{-3}$.

%Because the atomic and molecular gas do not mix in our simulation, we
%cannot produce cold atomic gas coinciding with molecular cloud
%cores. However, 
%EVS_rev1: I think we should not overemphasize this limitation. Besides,
%because there can be mixing at scales larger than a cell, it's not so bad.

With this information, we can observe cold atomic gas as it is about to
turn molecular. Even though atomic and molecular gas do not mix within
individual cells in our synthetic observations, it can still appear that
cold atomic gas coincides with molecular cores along a line of sight, if
the cold atomic gas is either directly in front or behind the molecular
gas. Because of this, our synthetic profiles should still look
realistic, as long as they include a significant number of grid cells,
which is generally the case. Note that, throughout this work, we use a
velocity range of -15 to 15 $\rm{km\ s^{-1}}$ and a velocity spacing of
0.15 $\rm{km\ s^{-1}}$.

For simplicity, we choose the line-of-sight direction along the
$x$-axis.  Due to the randomness of the initial condition, there is no
preferred direction and for our purposes it was unnecessary to
specifically select clouds and produce synthetic line profiles along
specific cloud axes. We leave those specific cuts for more specific
future applications.

%Compare: Glover et al. 2010 << too detailed and with evolving CO abundances and chemical modeling; 
%however, their Figure 5 shows that the mass-weighted CO abundance reaches 7-8 e-5 within a few Myr (cf. our 1e-4)
%They also find that H2 and CO form rapidly and that the limiting factor is building up sufficient densities
%also they find a poor correlation between CO and gas density!
The CO line profiles were derived under the assumption that the ratio of
the CO(1-0) to total molecular gas number density is (rather
approximately) $2 \times 10^{-4}$
\citep[e.g.][]{1984A&AS...58..327C,2008ApJ...679..481P}.
%\textbf{(why / refs? cf. Glover et al. 2010)}. 
%Watson & Salpeter 1972: [CO]/n ~2e-4 (so that's total hydrogen n?)
%Also Estalella & Anglada book p. 80 state [CO/H2] ~ 1.8e-4 without reference; and [12CO]/[13CO]=89
%Cernicharo & Bachiller 1984: [H2/13CO]=3.6e5
In our case, varying the exact CO abundance leads only to a scaling
effect in the line profiles, which in turn does not influence the
morphology of the gas overall. Since we have not computed any actual
chemical reactions, no further effects of the CO abundance exist in our
simulation.  

It is important to note that the critical density of molecular hydrogen
is determined by the local temperature. Thus, a drawback of our approach
is that we may miss some CO emission due to the limited resolution of
the gridded volume, implying that we may be smoothing out a potentially
clumpy medium. For example, if we have a molecular hydrogen density of
50 cm$^{-3}$ with our standard grid cell size of (0.5 pc)$^3$, and thus
would not be labeled as CO by our procedure. However, it is
possible that this gas was all concentrated in a volume of (0.23 pc)$^3$
in the original SPH data cube, thus having a density of 500 cm$^{-3}$,
which, depending on the temperature, would already be quite close to the
critical density for CO emission to occur, and therefore should be
labeled as CO.

We will distinguish between column densities derived from the simulation
directly and the inferred `observational' column densities. The former
are derived directly from the mass density cube by integration along the
LOS (which is 128 pc), assuming a mean particle mass of $\mu = 1.27$ for the calculation
of $N_{\rm{atom}}$, and $\mu = 2.36$ for $N_{\rm{mol}}$. On the other
hand, for 
%we directly calculated the atomic hydrogen column
%density $N_{\rm{atom}}$ and the molecular hydrogen column density
%$N_{\rm{mol}}$, assuming an average mass weight of 1.27 and 2.36 proton
%masses respectively. Separately, 
from the \HI\ and CO brightness
temperature profiles we also produce velocity-integrated maps of
$N_{\rm{H\sc{I}}}$ ($\rm{cm}^{-2}$) and $W_{\rm{CO}}$ ($\rm{K\ km\ s^{-1}}$)
respectively. These are the `observational' quantities. It should be
noted that $W_{\rm{CO}}$ is only calculated to provide a familiar
quantity and is not meant to predict any absolute values, as no distance
to the simulation volume was used or any kind of telescope resolution
effects. The \HI\ column density was computed from the \HI\ brightness
temperature profile using the common assumption of the atomic gas being
optically thin.

\subsection{Analysis of the profiles} \label{sec:profile_analysis}

\begin{figure}
  \resizebox{\columnwidth}{!}{\includegraphics{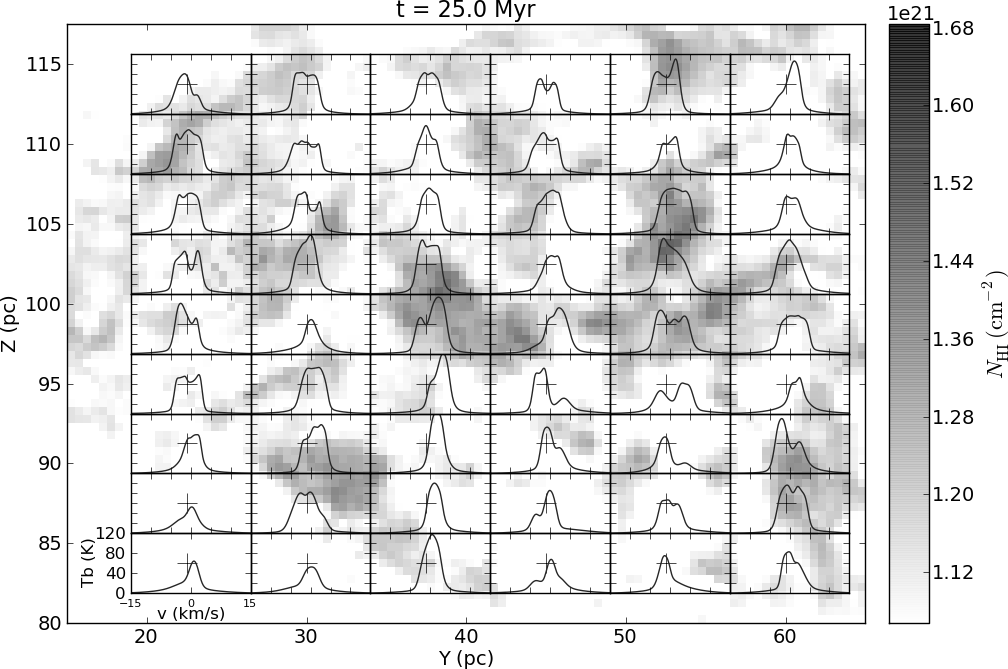}}
  \resizebox{\columnwidth}{!}{\includegraphics{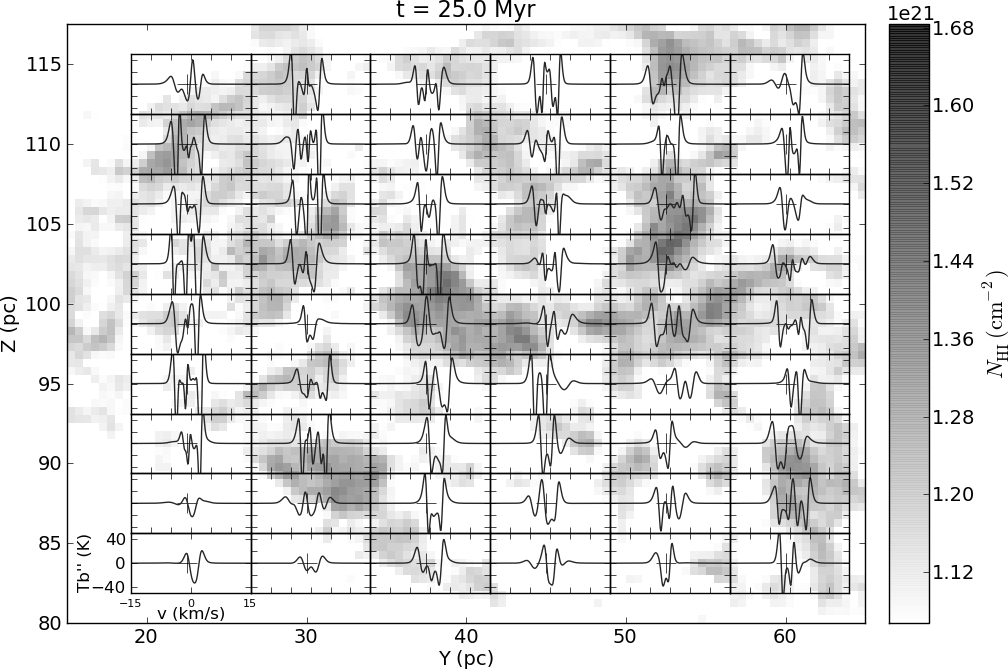}}
  \caption{\label{fig:NHI_insets} {\it Top panel}: Detail of the atomic
  hydrogen column density map produced with our simple radiative
  transfer code. Brightness temperature profiles are overlayed to show
  the velocity distributions at the locations marked with
  plus-signs. It can be seen that most
  profiles show central dips, particularly where the \HI\ column density
  is higher. {\it Bottom panel}: The second derivative of the brightness
  temperature profile is shown for the same area. Dips in the brightness
  temperature profile are relatively easy to detect as peaks in the
  second derivative.}
\end{figure}

Figure \ref{fig:NHI_Nmol_select} shows the integrated \HI\ column
densities at our three timesteps and selection boxes (white rectangles)
highlighting 
regions of interest because of the presence of molecular cloud
structure.  Extended gas structures are visible at a level of about 30
cm$^{-3}$, permeating the entire 128-pc gridded volume. Our
volume contains a large underdense region (below 1 cm$^{-3}$) that
persists throughout our three timesteps under consideration, surrounding
the dense, evolving dense clouds.

It is remarkable that, in spite of the obvious constraints and
limitations of the simulation, there is a striking resemblance to the
Taurus molecular cloud and its immediate surroundings in the first
timestep (Figure \ref{fig:NHI_Nmol_select}). The region contains a
molecular cloud in the upper left hand side, next to something that
appears to be a cavity in the \HI\ gas, surrounded by a diagonal band of
\HI. This type of structure could be labeled as an \HI\ (super-)shell,
although our simulations does not include the kind of supernova feedback
that might produce \HI\ supershells. We merely point out the resemblance
to caution that the 2D projection of the atomic gas may look like a
supershell, but further kinematic confirmation should be sought.

Figure \ref{fig:NHI_insets} shows a zoom into the selection box of the
25-Myr timestep, showing 
%an example region where
%we zoomed in on a concentration of atomic gas. For a regularly spaced 
%grid, we show 
\HI\ brightness temperature profiles at various locations.
% (replacing the x-direction). 
The \HI\ profiles almost universally show a depression around the
zero-offset velocity. This depression may be caused either by a lack of
atomic gas or by self-absorption. It is hard to determine from the
observations alone which of these possibilities is the actual cause of
the dip in the profile, but in the rest of this contribution we will
explore the possibility of doing so, taking advantage of our having the
full information about the fluid variables available in our simulated volume.

%Shetty et al. 2011 use RADMC-3D (Dullemond et al.) for RT. Extensive descriptions with formulas.

\begin{figure*}
  \resizebox{0.8\columnwidth}{!}{\includegraphics{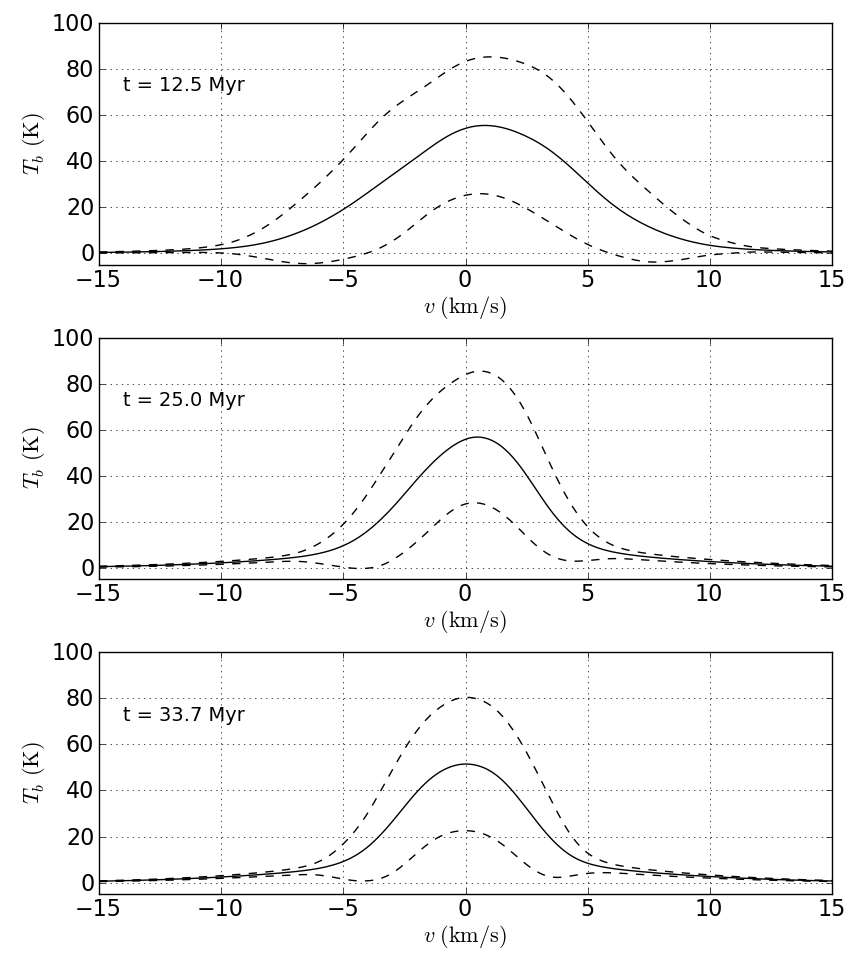}}
  \resizebox{0.8\columnwidth}{!}{\includegraphics{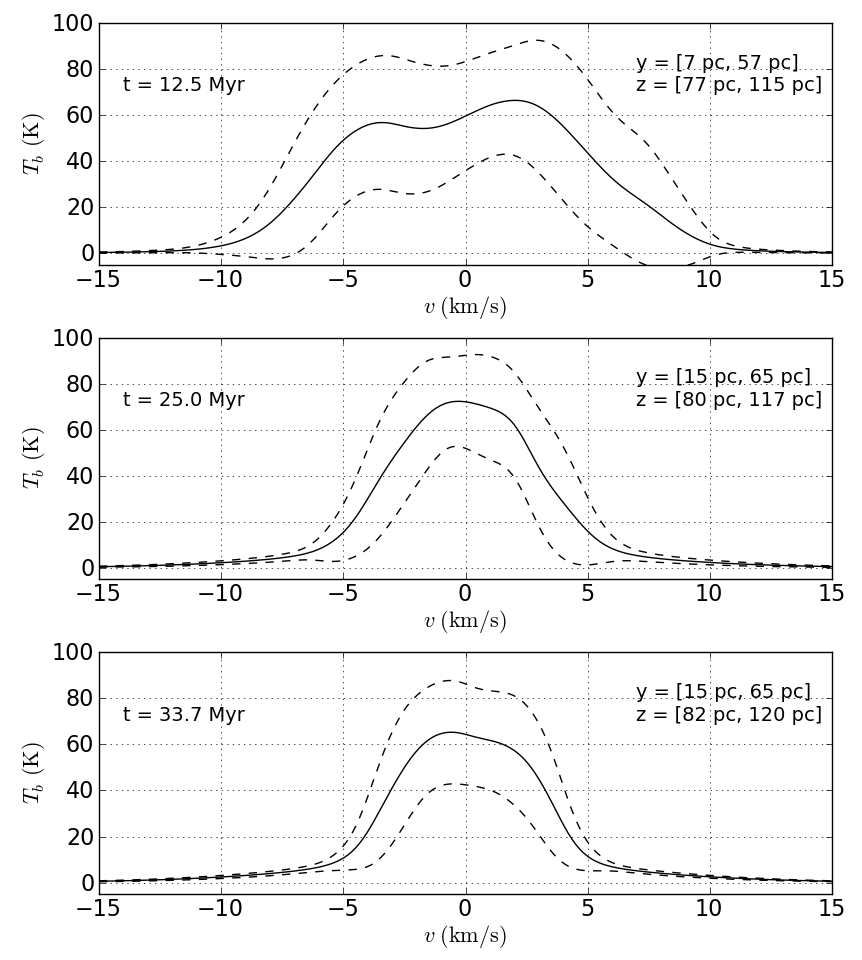}}
  \caption{\label{fig:avg} {\it Left column}: Averaged brightness temperature
  profiles of the full gridded cubes are displayed with one-sigma
  deviations from the average value as dashed lines. It can be seen that
  the velocity dispersion is becoming lower at later times. {\it Right
  column}: The same averaged temperature profiles with one-sigma
  deviations are shown as in the previous figure, but for a specific
  region of the simulation box. More structure in the profiles can be
  seen in this case; in particular, dips.}
\end{figure*}

\begin{figure}
  \resizebox{\columnwidth}{!}{\includegraphics{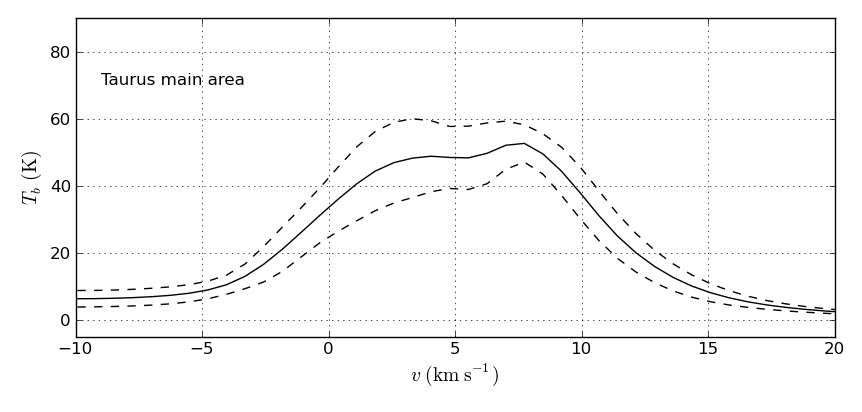}}
  \caption{\label{fig:avg_taurus} Averaged \HI\ brightness temperature
  profile ({\it solid line}) for the main region of Taurus, with
  one-sigma deviations ({\it dashed lines}). The broad profile with a
  central dip is typical for line profiles in the Taurus region.}
\end{figure}

Averaging all \HI\ brightness profiles in the gridded cube leads to
relatively featureless profiles (Figure \ref{fig:avg}, left column),
although they 
%these profiles (one per timestep) 
do show that the velocity dispersion of the gas in the gridded cubes
decreases as time progresses, due to the decaying turbulence. On the
other hand, zooming into a specific region of the simulation box where
molecular gas has formed yields profiles with more structure. Most
notably, the profiles display the familiar central depression (Figure
\ref{fig:avg}, right column).

Since the smaller region is selected to be approximately at the same
spatial location at all three times shown, the averaged profile changes
significantly from one time to the other. At the middle time, very
little structure is present, while at the other timesteps more structure
is seen, which is most likely caused by converging flows of
gas. Finally, to show the qualitative agreement, we show in Figure
\ref{fig:avg_taurus} an averaged \HI\ profile of the main Taurus region,
using data from the Arecibo GALFA-\HI\ \citep{2011ApJS..194...20P}
survey. Particularly, the velocity dispersion is comparable. The Taurus cloud is estimated to have linear dimensions of 32
$\times$ 5 pc \citep[e.g.,][] {2009MNRAS.395L..81B}. Our simulation
volume is large enough to contain Taurus-like structures. The velocity
dispersion at the earlier time shown from our simulation is the closest
to what is seen in Taurus, consistent with the view that the Taurus area
is still relatively young \citep[e.g.,][] {2001ApJ...562..852H,
2003ApJ...590..348L, 2007ApJ...659.1629L, 2010ApJ...721..686P}.

\section{Molecular gas and HISA} \label{sec:HISA_molec}

\subsection{Identifying \HI\ self-absorption features}

%It would be interesting to provide some statistics on our HI(N)SA, such as:
%Profiles above a certain bg (not many, presumably)
%Notably, dips steeper than shoulders (based on first derivative)
%Other?

\begin{figure}
  \resizebox{\columnwidth}{!}{\includegraphics{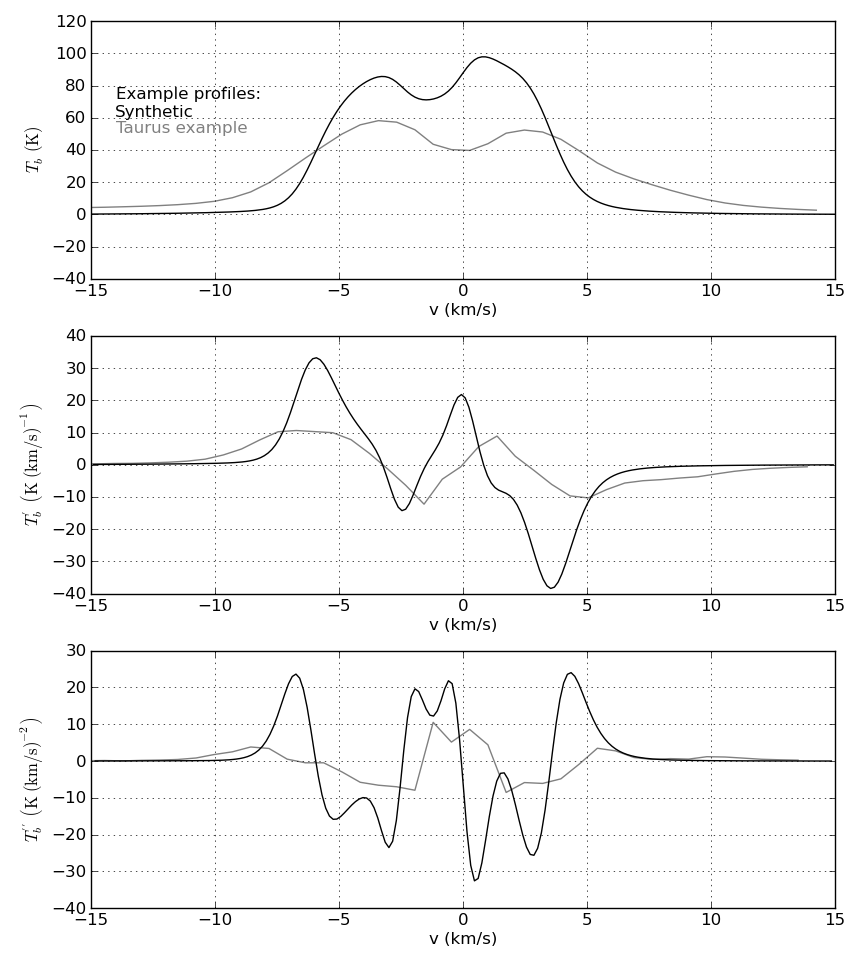}}
  \caption{\label{fig:compare_profile} {\it Top panel:} Typical
  atomic hydrogen line profile of the Taurus molecular cloud, 
  exhibiting double-peaked brightness temperature distribution
  (gray). The black line shows a similar line profile from our
  simulation. {\it Middle and bottom panels:} first and second
  derivatives of the line profiles, respectively. 
  Note that we chose a velocity resolution in the synthetic
  observations that is several times better than the Taurus observations
  that we used, namely 0.15 and 0.74 km s$^{-1}$ respectively.
%The second derivative
%  can be used
%  in principle to trace \HI\ (narrow) self-absorption but can be
%  confused with colliding inflows of gas. {\bf JONATHAN: Why do you say
%  this?}
  }
\end{figure}

\begin{figure*}
  \resizebox{1.8\columnwidth}{!}{\includegraphics{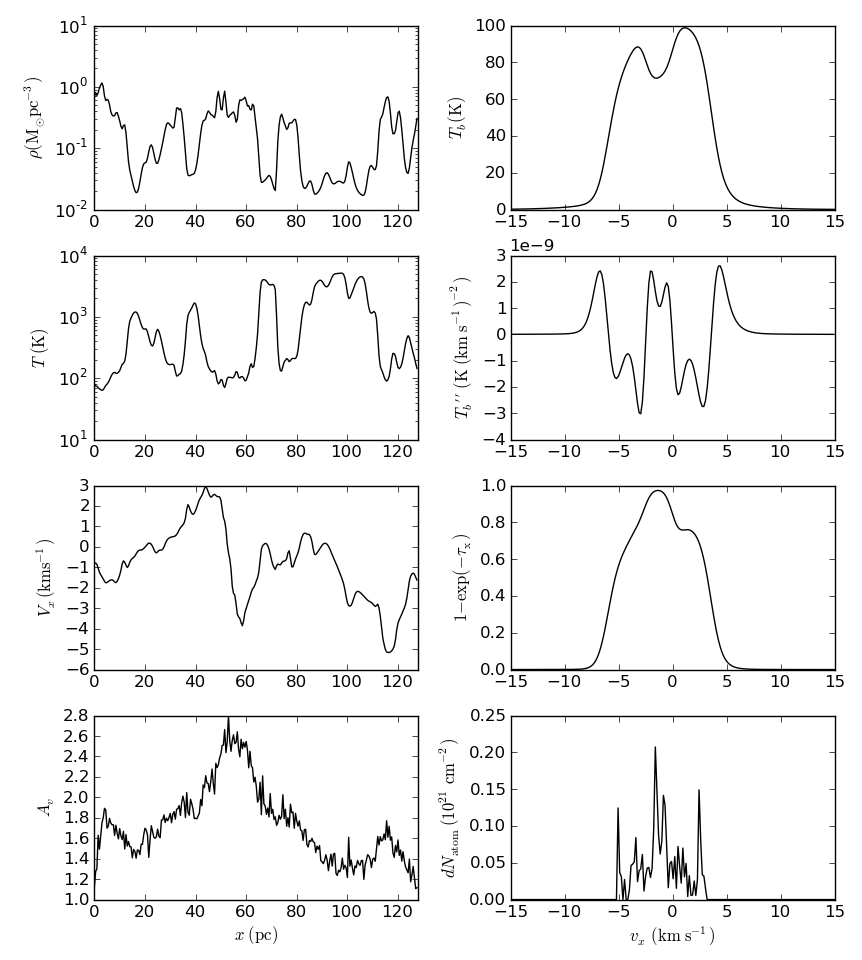}}
  \caption{\label{fig:multiplot} Plots showing the profile from Figure
  \ref{fig:compare_profile} with various quantities to illustrate the
  structure behind the \HI\ profile. The quantities shown are: Total gas
  density $\rho$, $T$, $v_x$, and $A_v$, as well as the \HI\ brightness
  temperature and its second derivative, the opacity $\tau_x$ (plotted
  as the extinction (1 - exp(-$\tau$))) and the fractional
  $N_{\rm{atom}}$.}
\end{figure*}

In \citet{2013MNRAS.429.3584H}, we used the second derivative of the \HI\ 
profile as a tracer for \HI\ self-absorption, and by extension, as a
tracer for cold atomic gas. This approach was inspired by
\citet{2008ApJ...689..276K}, who used the second derivative of the \HI\
line profile in combination with molecular (CO) emission line data to
correct the \HI\ profiles for self-absorption. Their rationale was that
\HI\ line profiles are approximately gaussian in nature, and that
the profiles of HISA are much narrower than those of the background
emission, so that
%subsequent derivatives of {\bf the total} profile will
%have a rapidly increasing dependence on the linewidth of the profile: 
an \HI\ self-absorption feature will become progressively more dominant
in progressively higher derivative profiles, while the overall emission
profile will fade into the background. \citet{2008ApJ...689..276K} 
concluded
that using the second derivative is sufficient to locate potential
self-absorption features with relative ease. Additionally, they required
the presence of CO emission: When it was present, they referred to this
as `\HI\ narrow self-absorption', or HINSA for short, although we
%, as opposed to HISA. We 
will stick to the more general term HISA here.
%, since CO is not
%necessarily present at locations where we determine \HI\ self-absorption
%to be occurring.
%EVS_rev1: I commented out the next two sentences because I feel they're 
%not essential, and they distract from the main line of reasoning.
%
%It should also be noted that in
%\citet{2013MNRAS.429.3584H}, we simply used the amplitude of the second
%derivative of the \HI\ profile to infer the presence of molecular gas
%from the \HI\ data for Taurus,
%and not the full procedure of
%\citet{2008ApJ...689..276K}, which includes deriving a
%best-fit opacity and an inferred cold gas column.  
%Whether the molecular
%gas was actually there at the same location (Taurus) was of lesser
%importance since the only condition that needed to be satisfied was that
%\HI\ produced by photodissociation or in the process of being converted to
%molecular hydrogen was present. 
In what follows, we will explore the uncertainties
involved in using the second derivative of the \HI\ profile, whether or
not supplemented by CO observations.

In the top panel of Figure \ref{fig:compare_profile} we show typical
\HI\ line profiles, and
their first and second derivatives respectively in the middle and bottom
panels. The black lines represent a profile from our simulation at 12.5 Myr 
at an arbitrary location, whereas the gray line
represents a typical \HI\ profile from the Taurus molecular cloud, from
the GALFA survey. 
%The dash-dotted lines represent fits of three
%gaussians to the profile, whereas the dashed lines show the combined
%fit. 
It can indeed be seen that the dips in the profiles become prominent
in the second derivative, %({\bf Jonathan: to my eyes, the central feature
%is not more prominent than the wings, no?}). 
since sharper dips produce stronger features in the second derivative. 
The Taurus profile is starting to be dominated by noise in the second
derivative, but its HISA-indicating peak is still prominent.
Note that the wings of the second derivative of the profile in the synthetic
case are much more prominent, which is a result of our brightness temperature profiles being
steeper than a typical Taurus profile. This property is an important
clue to the composition and physical conditions of the gas, but a
more thorough investigation of 
this issue
%particular subject 
is beyond the immediate scope of this work.

\begin{figure*}
  \resizebox{0.65\columnwidth}{!}{\includegraphics{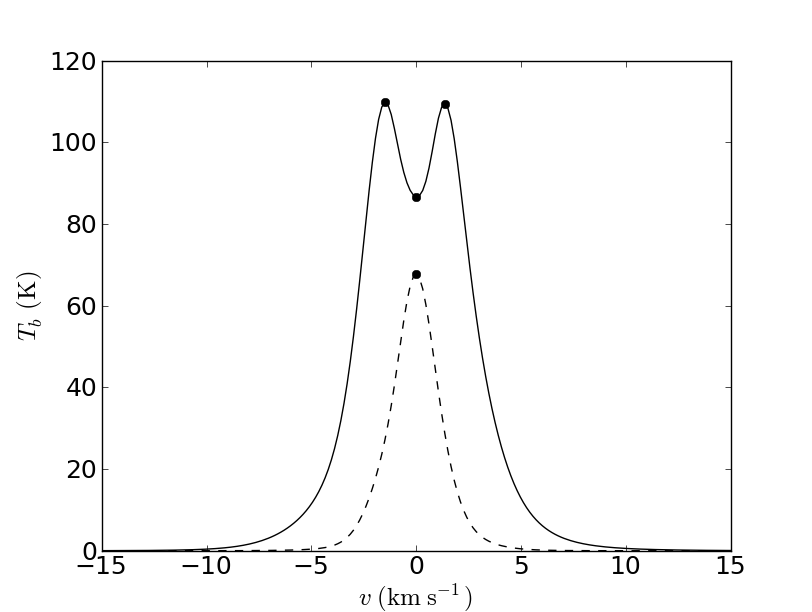}}
  \resizebox{0.65\columnwidth}{!}{\includegraphics{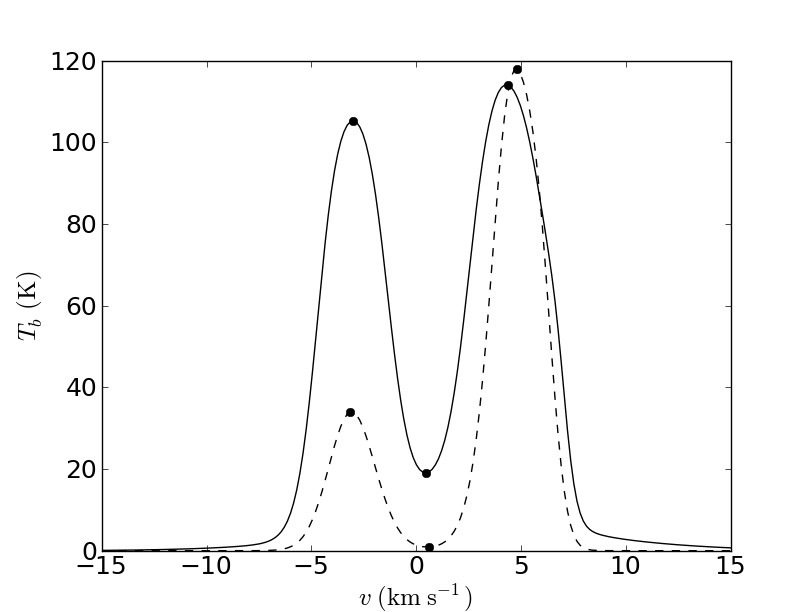}}
  \resizebox{0.65\columnwidth}{!}{\includegraphics{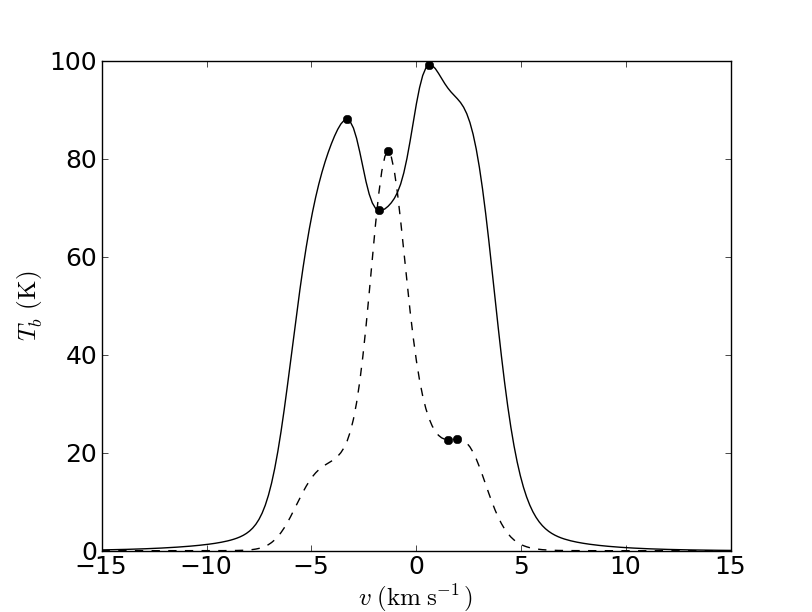}}
  \caption{\label{fig:hinsamatch} We verify the presence of HISA by
  matching local minima in the \HI\ brightness temperature profile
  (solid lines) to peaks in the optical depth $\tau$ (dotted lines) along the line of sight. The
  value of $\tau$ has been exaggerated in these plots for clarity and is
  shown at an arbitrary scale. {\it Left panel}: A simple profile with a dip
  due to self-absorption. {\it Middle panel}: No HISA is present since the dip is shown
  to be due to a lack of gas, not due to extinction. {\it Right panel}: A more
  elaborate profile with a HISA dip and with a component on the right
  hand side that is not detected as HISA in our procedure but is in fact
  due to absorption. All profiles were extracted from our simulation.}
\end{figure*}

As mentioned above, it is often unclear from observations of the \HI\
line profile whether a dip in the profile is caused by self-absorption
or by a lack of atomic gas. Depending on the local morphology of the
cloud, the lack of atomic gas may imply the presence of molecular
gas, but then an additional tracer of the molecules is needed.  
%EVS_rev1: Commented out the following statement to avoid repetition.
%The
%advantage of dealing with simulations is that in principle all the
%needed information is available. For example, 
Figure \ref{fig:multiplot} shows, for the position on the $(y,z)$ plane
used in Figure \ref{fig:compare_profile}, the density, temperature,
velocity distribution and visual extinction along the $x$ spatial coordinate, as well as the brightness temperature, its second derivative,
opacity and atomic column density per velocity interval (in the
velocity coordinate).
\citet{2007A&A...465..445H} showed similar plots and noted that individual
cloudlets cannot be distinguished observationally along the line of sight
due to thermal broadening. The same can be seen in our figure.

Since our goal is to test how well HISA correlates with the molecular
emission, we need to make sure that what we identify as HISA is indeed
due to \HI\ self-absorption, and not just to a lack of emission. To this
end, we take advantage of the fact that the radiative transfer code
producing the \HI\ line profile also records the local opacity. Thus, we
can easily search for those instances where dips in the \HI\ profile
closely match peaks in the opacity ($\tau_x$) along the velocity
coordinate. 

Figure \ref{fig:hinsamatch} illustrates this process 
for a profile with an emission dip matched by an opacity peak
(left panel), a non-HISA profile (middle panel), and a slightly
more complex profile (right panel) --- all taken from individual
positions in the simulation. The peak in the opacity does not have to be
exactly at the same velocity as the minimum brightness temperature of
the \HI\ profile dip, as there could be foreground or background
material emitting in the same velocity interval, partially filling up
the dip and displacing it from the opacity peak. We thus (arbitrarily) choose
to find opacity peaks in a range of $\pm 0.6\ \rm{km\ s^{-1}}$, or four
velocity channels in each direction. 

Not all instances of HISA are
located this way, though, as the profile on the right panel of Fig.\
\ref{fig:hinsamatch} illustrates. The second opacity 
peak in this plot results in an inflexion point in the \HI\ profile,
which we do not count as HISA, since we only flag local minima. But at
least we guarantee that everything that we label as HISA indeed
corresponds to self-absorption.  Additionally, we record the depth of
the HISA-induced feature relative to the adjacent peaks in the profile,
a quantity we refer to as the {\it dip depth}. We use this
information to check for correlations between the depth of the HISA
features and the local molecular gas content. 
%, since the presence of cold
%atomic gas may signal that molecular gas is in the process of forming
%nearby, or was formed recently.

Applying this \HI-minimum/opacity-peak matching process for every
spatial pixel, we obtain a mask of locations where true HISA is
detected.  We will refer to this as the {\it HISA mask map}. In
addition to being a mask, however, this map contains the values of the
HISA dip depth, thus providing not only an on/off switch for where
`true HISA' is detected, but also a
measure of the strength of the absorption feature. When we consider
these values for comparison, we will indistinctly refer to this as the
`HISA dip depth map'. To reduce the
number of false positives, we also require that the matching opacity
peak be closer to the dip in $T_b$ than the nearest opacity (local)
minimum. This mask can then be compared to the HISA derived from the
second-derivative method and to molecular gas emission.
%It should also be stressed once again
%that our simulation does not allow mixing of cold atomic and molecular
%gas. Therefore, by design, we miss a very specific population of
%HISA-producing gas.  

\subsection{Strength of the HISA features compared to the `real' HISA}

\begin{figure*}
  \resizebox{2\columnwidth}{!}{\includegraphics{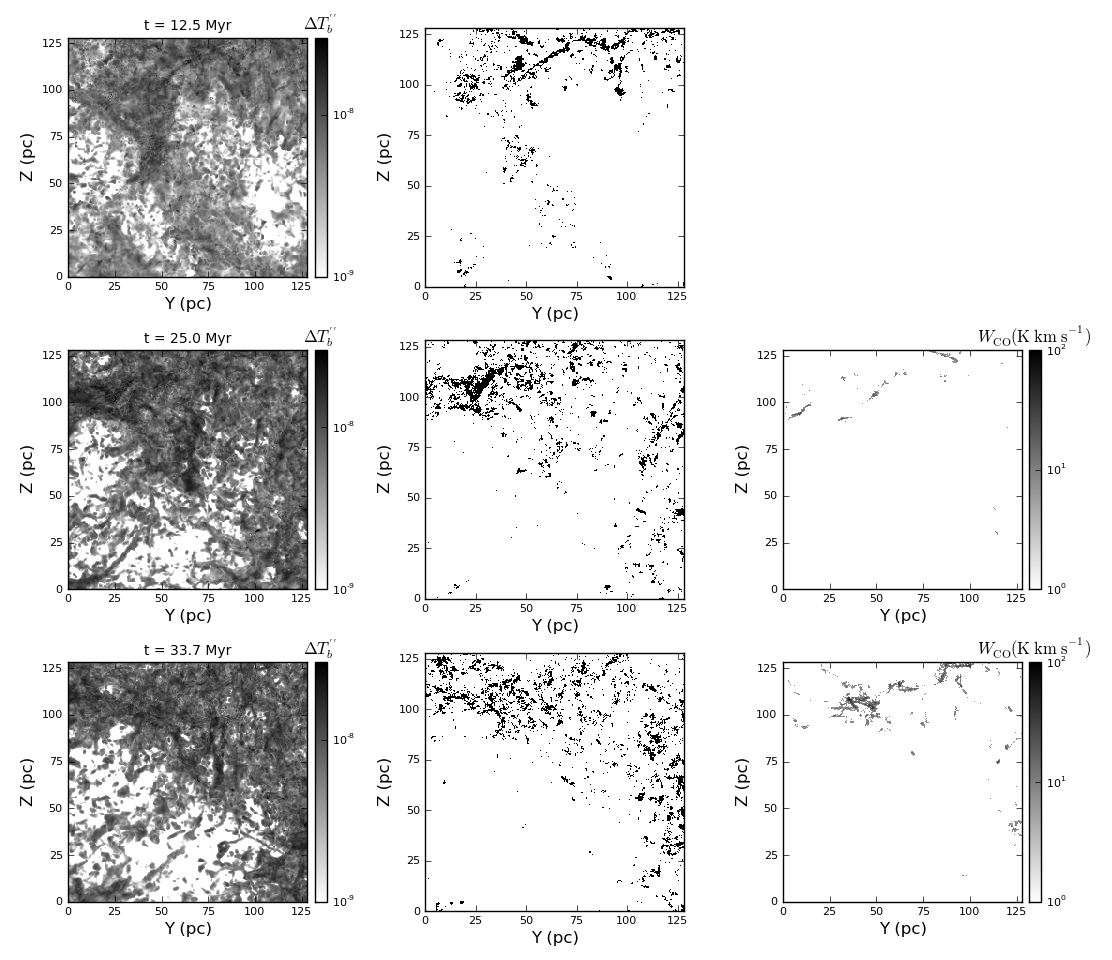}}
  \caption{\label{fig:mm_mask_WCO} {\it Left column:} HISA strength
  maps at the three timesteps we study (with time increasing from top to
  bottom), where the HISA strength is defined as the maximum difference
  between values of the second derivative of the brightness temperature
  profile as in \citet{2013MNRAS.429.3584H}. {\it Middle column:} the HISA mask, or `real HISA', defined
  as the set of pixels in a projection of the simulation where local
  minima in the \HI\ line profile coincide with peaks in the optical
  depth. While the major cloud structures are visible in
  both the left and middle columns, the correspondence is far from
  perfect. {\it Right column:} Integrated CO intensity maps are shown
  for the two timesteps where CO emission is present.}
\end{figure*}

Analogously to the procedure inspired by \citet{2008ApJ...689..276K}
that we used in \citet{2013MNRAS.429.3584H}, we have constructed a 'HISA
strength' map that is simply the maximum minus the minimum value of the
second derivative of the \HI\ brightness temperature profile 
($T^{\prime\prime}_{\rm{max}} - T^{\prime\prime}_{\rm{min}}$). Since
narrow self-absorption starts to feature prominently in the second
derivative of the line profile, measuring the largest local
difference in its value can be considered a reasonable proxy for the
presence and the amplitude of the HISA feature. However, based on the
\HI\ profile alone one cannot distinguish between HISA and the lack of
atomic gas, which is why
\citet{2008ApJ...689..276K} required the presence CO emission at the
position on the plane of the sky (POS) where HISA was to be identified,
as a condition to declare the reduced \HI\ emission as indicative of
HISA. In \citet{2013MNRAS.429.3584H}, we did not explicitly require CO
emission, but the region of interest (Taurus) is a well-known molecular
cloud, with a well-known presence of CO, so dropping the CO requirement
had no obvious consequences. Now, we can begin to use our simulation to
estimate how reliably the HISA strength map, by itself, traces HISA.
%, while
%remembering the approximation of not having atomic and molecular gas at
%the same location. 

Figure
\ref{fig:mm_mask_WCO} shows, from left to right, the HISA strength map,
the HISA mask and the integrated CO intensity. 
It can be seen that the HISA strength map shows the most features,
which in part can be seen on the HISA mask. Finally, only a small fraction
of the features in either map are seen in CO emission.

%easily that qualitatively
%there is a correlation, especially 
%between the HISA mask and the
%molecular hydrogen column density map, but at our resolution of 0.5
%parsecs, the correlation is weak.

\begin{figure}
  \resizebox{\columnwidth}{!}{\includegraphics{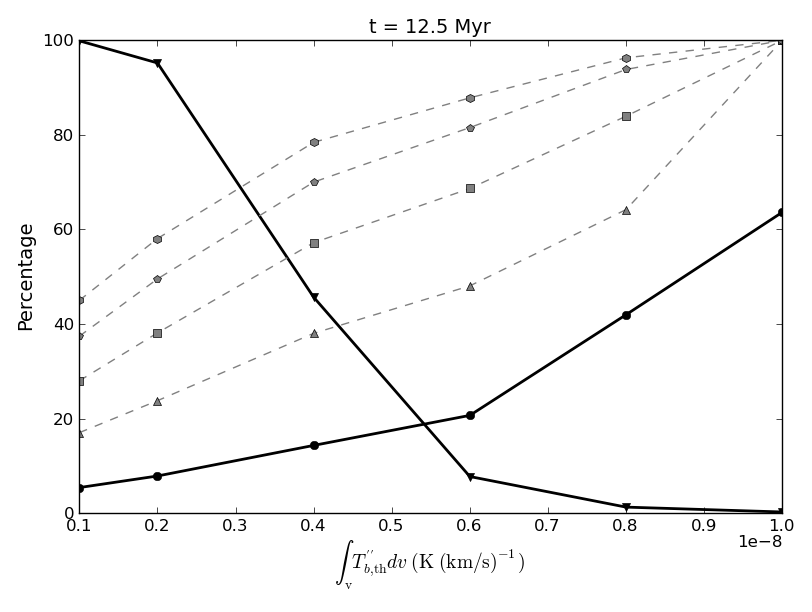}}
  \resizebox{\columnwidth}{!}{\includegraphics{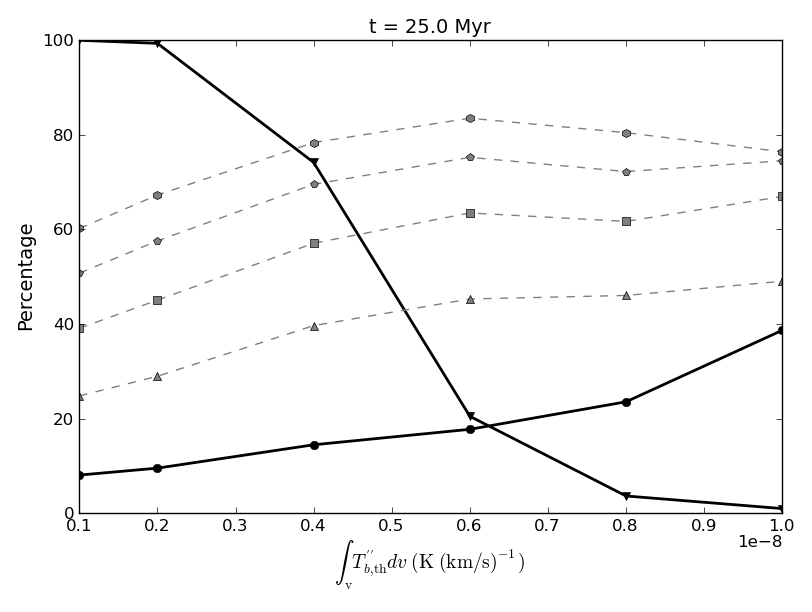}}
  \resizebox{\columnwidth}{!}{\includegraphics{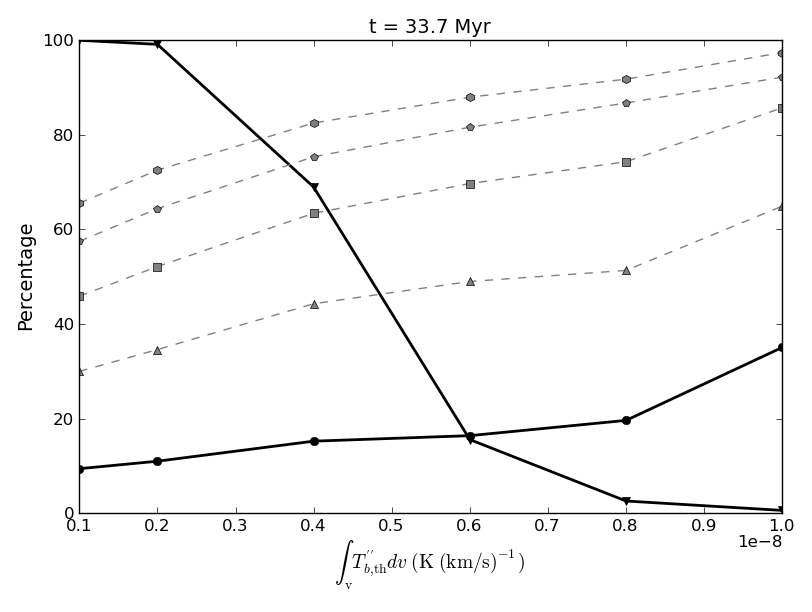}}
  \caption{\label{fig:successrate} `Match success rate', or
  fraction of pixels in the HISA strength map ({\it left column} of
  Fig.\ \ref{fig:mm_mask_WCO}) above a certain threshold value that
  coincide with the HISA mask ({\it middle column} of Fig.\
  \ref{fig:mm_mask_WCO}), shown percentage-wise, as a function of the
  threshold value, for the three timesteps studied (solid line with
  circular dots). The dashed grey lines with upright triangles,
  squares, pentagons and hexagons represent the match success rate
  with pixels within distances of 1, 2, 3, and 4 pixels,
  respectively. The solid line with upside-down triangular markers shows
  the number of pixels above the threshold value as a percentage of the
  number of pixels in the HISA mask. }
\end{figure}

\begin{figure}
  \resizebox{\columnwidth}{!}{\includegraphics{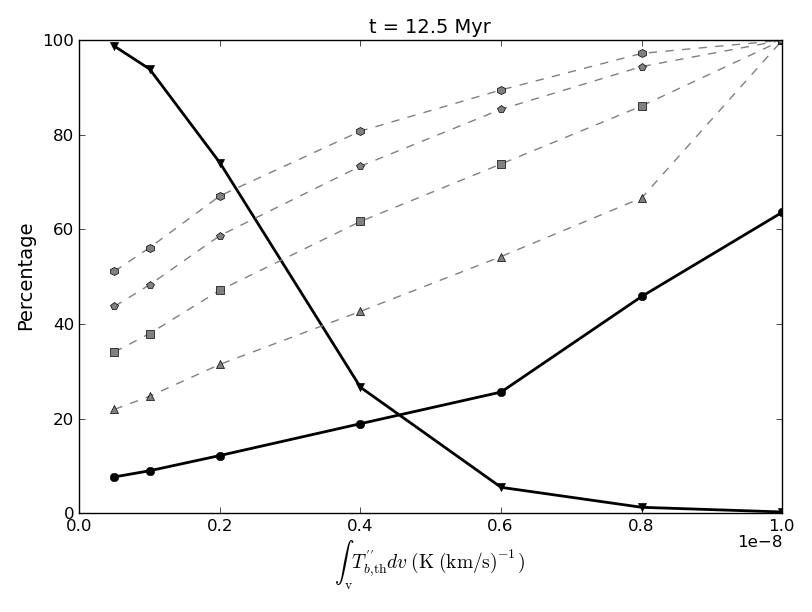}}
  \resizebox{\columnwidth}{!}{\includegraphics{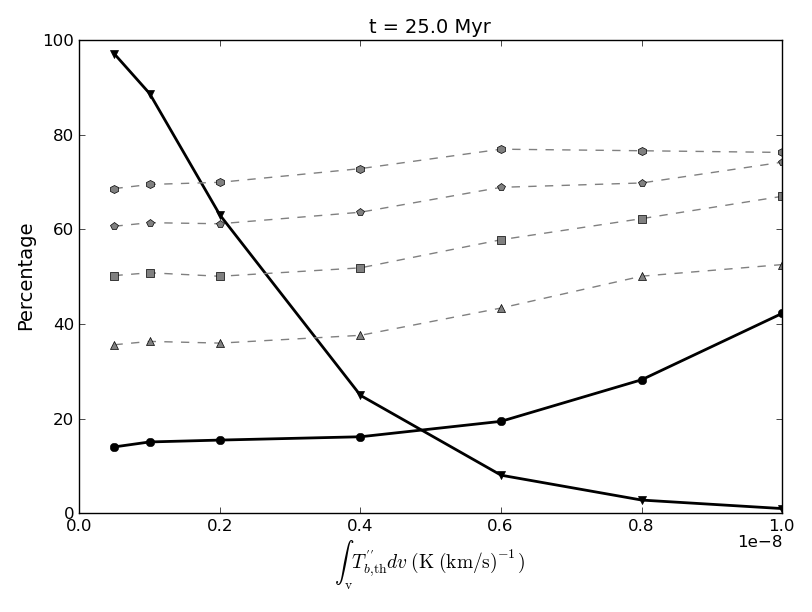}}
  \resizebox{\columnwidth}{!}{\includegraphics{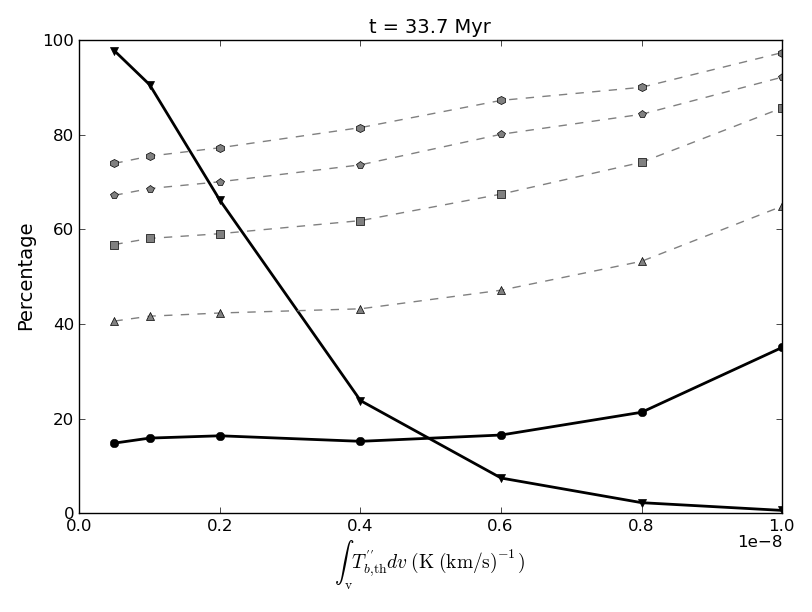}}
  \caption{\label{fig:successrate_blank} The same quantities are plotted
  as in Figure \ref{fig:successrate}, but calculated from a HISA
  strength map with the wings on both ends of $T_b^{\prime\prime}$
  blanked out. No significant improvement in the success rate was
  obtained by applying the wing blanketing.}
\end{figure}

By using the term HISA \textit{strength}, we assume (or imply) that
higher values of HISA strength correspond to higher odds of a particular
pixel in the map representing actual HISA. In reality, there is no clear
relation between the HISA strength and any actual HISA amplitude. Still,
we test this assumption in Figure \ref{fig:successrate}, by showing how
well the pixels in the HISA strength map correspond with the HISA
mask. 
%While not perfect, the HISA 
%mask strength map (JONATHAN: correct? The original text said ``HISA
%mask'', but I think the mask is the real HISA, and the strength is the
%observational proxy.) does in fact represent actual HISA as it is found
%in the simulation.

In order to make this comparison,
% we have to adopt a slightly different
%definition of HISA strength. While so far we have used it to refer to
%the amplitude of the second derivative of the \HI\ brightness temperature
%by calculating the distance between the maximum and the minimum value of
%the second derivative, now we will 
we adopted a slightly different definition of HISA strength that allows
different thresholding values, namely, 
we used increasing threshold values of the integrated second
derivative of the \HI\ profile, summing the contributions of the points
along the LOS that are above the threshold. %{\bf (JONATHAN: Did you
%really use a different operational definition for the HISA strength here
%than previously defined? If so, I see no point in having two different
%definitions for the same quantity, and I suggest to use just one --
%whichever you think is more useful -- throughout the paper.)}
% for the second derivative and integrate over velocity channels where the
%value of the second derivative is above the threshold value. This is
%comparable to the minimum/maximum definition of HISA strength, but
This allows us to consider a scale of increasing HISA strength. As
this threshold is increased, fewer and fewer pixels will be captured in
the spatial plane. Thus, we only count the number of pixels that are
left after applying each threshold value, and then check how many of
those pixels also lie on the HISA mask.

% ({\bf
%Also, doesn't the fact that we are adding up the points where the 2nd
%derivative is larger than the threshold along the LOS mean that we may
%be getting false positives if two independent clumps are aligned along
%the LOS?  <<not what we did
%This in reality does not produce a stronger absorption feature,
%but it would be selected by the summation of the second derivative
%amplitudes, no? I'm thinking that perhaps what we should as the proxy
%for HISA, is just the maximum value of the derivative along the LOS, no? 
%What do Krco et al actually do?

%Finally, I notice that the wings are a lot more prominent on the 2nd
%derivative for the synthetic data than for the Taurus profile. This may
%have consequences, as I mention below.})

We define a `match success rate' as the percentage of pixels
where the HISA strength pixels above a certain threshold match
with corresponding HISA mask pixels (implicitly assuming the HISA mask
to be superior, although not perfect). We also counted the percentage of
pixels in the HISA mask map that are above the threshold in the
HISA strength map.
% relative to the number of pixels in the HISA mask.
%It 
This latter percentage drops monotonically for increasing values
of the threshold, reflecting the declining number of pixels in the HISA
strength map as the threshold value increases. Combining the two
percentages (Fig.\ \ref{fig:successrate}) shows that while the
success rate goes up for increasing values of the threshold, the number
of points with which this is achieved goes down. For lack of a perfect
correspondence, a balance should be achieved between the success rate
and the number of pixels with which this is achieved if one were to optimize the detection of HISA. 
%A 100\% success
%rate carries no significance if it is achieved with only one pixel, for
%example.

Since we already noted that the HISA strength map and the HISA mask
agree 
%at least on a qualitative level, 
well on a coarse scale, but less accurately on a fine scale (cf.\
Fig.\ \ref{fig:mm_mask_WCO}), we also check for nearby HISA mask
pixels for every HISA strength map pixel, which is effectively a form of
smoothing. Figure \ref{fig:successrate} also shows the success
rates when HISA mask pixels are found within one, two and three pixels
%range 
respectively, where by `a distance of one pixel' we mean that we
search in a $3 \times 3$ -pixel box (or, equivalently, in a $1.5
\times 1.5$-pc box) with the HISA strength map
pixel at the center. Interestingly, even searching within 
just one pixel already doubles the typical success rate, while 
%indicating that the maps are indeed qualitatively similar. Searching 
searching within three
pixels (a $3.5 \times 3.5$-pc box) approximately quadruples the
original success rate or more. This means that, while the HISA
strength maps do not correspond very well
%in detail 
to the occurrence of actual HISA (as
approximated by the HISA 
%match
 mask map) on a pixel-by-pixel basis, they do correspond in
majority to actual HISA 
%on 
 within a 
%scale 
distance of about 3 pc.

Comparing the three different timesteps, the first thing to note is the
sharp drop in the percentage reflecting the number of pixels above the
threshold values. At the highest threshold value, this percentage
reaches a nearly zero value in all three timesteps, indicating that
applying a higher threshold will yield no improvement. Next, for the
first timestep we see that the success rate increases sharply with an
increasing threshold, although the number of pixels in the mask drops
sharply as well. However, for the other two timesteps the success rate
does not rise as sharply with an increasing threshold, even when smoothing is applied. Basically, in
%this case 
these cases, increasing the threshold value does not 
significantly improve the correspondence, and the only way to improve it is to apply smoothing.

One potential problem with using the second derivative of the brightness
temperature profile ($T_b$) is that in our simulations, the wings of the
profile become very prominent in the 
%of 
$T_b^{\prime\prime}$ 
%are very dominant
profile, which is indicative of 
%a 
steep line profile wings (see e.g. Figure
\ref{fig:compare_profile}). In the $T_b^{\prime\prime}$ profile,
these wings  
can have higher amplitudes than the peak caused by HISA at the center of
the profile. Because of our definitions of HISA strength that are
directly related to the amplitude of $T_b^{\prime\prime}$, it
is possible that our HISA strength measures are contaminated by
$T_b$ profiles that are purely steep with no absorption feature.
By comparison, the Taurus line profiles have relatively broad and
shallow wings and
therefore do not suffer from this problem: the peak in $T_b^{\prime\prime}$ is
caused almost exclusively due to dips in the center of the $T_b$ profile.
However, this could be due partially to the poorer velocity resolution of the
Taurus data.

In order to see whether this contamination influenced our success rate,
we blindly blanketed out (i.e. set $T_b^{\prime\prime}$ to 0) the outer
wings of the second derivative. This is fairly trivial since all emission
profiles increase initially from both ends of the velocity axis towards a peak.
The resulting success rate plots are shown in Figure \ref{fig:successrate_blank},
where it can be seen that this procedure has very little if any effect on
the success rate. In other words, the steep profiles and the resulting
dominant wings in the second derivative of the line profile do not cause
significant problems in the identification of HISA features.

%({\bf JONATHAN: now that I think of this, I'm a bit skeptical. Could
%this be due to the fact that we are adding the points above the
%threshold rather than taking the maximum value of the 2nd derivative?
%Since the wings are very prominent in the 2nd derivative map for the
%synthetic observations, it could be that we are sometimes picking up
%points that don't have a very strong absorption, but appear in the HISA
%strength map because they have a strong contribution from the
%wings. Could you check by just taking the maximum value of the 2nd
%derivative, rather than the sum of the points above the threshold?})

%\textit{Reiterate that we do not form cold atomic gas in the same volume as molecular gas! Therefore our results could be pessimistic.}

%\textbf{\textit{Edits until here! (HI)}}

\subsection{Comparing HISA to the CO emission}

%{\bf JONATHAN: I think the discussion in this section mixes up the CO
%and CO-free emissions. We need to compare each one separately
%with the HISA, and then compare the two among themselves.}

% \begin{figure*}
%   \resizebox{0.6\columnwidth}{!}{\includegraphics{Figures/t096_NHI_WCO_Hd}}
%   \resizebox{0.6\columnwidth}{!}{\includegraphics{Figures/t192_NHI_WCO_Hd}}
%   \resizebox{0.6\columnwidth}{!}{\includegraphics{Figures/t259_NHI_WCO_Hd}}
%   \caption{\label{fig:hinsadepth} We show qualitative images of the spatial distribution of the atomic hydrogen column density (blue), the integrated CO intensity (green) and the depth of the dip caused by HISA (red).}
% \end{figure*}

% \begin{figure*}
%   \resizebox{0.6\columnwidth}{!}{\includegraphics{Figures/t096_NHI_WCO_Hd_zoom}}
%   \resizebox{0.6\columnwidth}{!}{\includegraphics{Figures/t192_NHI_WCO_Hd_zoom}}
%   \resizebox{0.6\columnwidth}{!}{\includegraphics{Figures/t259_NHI_WCO_Hd_zoom}}
%   \caption{\label{fig:hinsadepthzoom} Here we show similarly qualitative images like the ones in Figure \ref{fig:hinsadepth}, but zoomed to show in more detail the location of HISA features (red) relative to the CO emission (green) and the surrounding \HI\ gas (blue).}
% \end{figure*}

\begin{figure*}
  \resizebox{2\columnwidth}{!}{\includegraphics{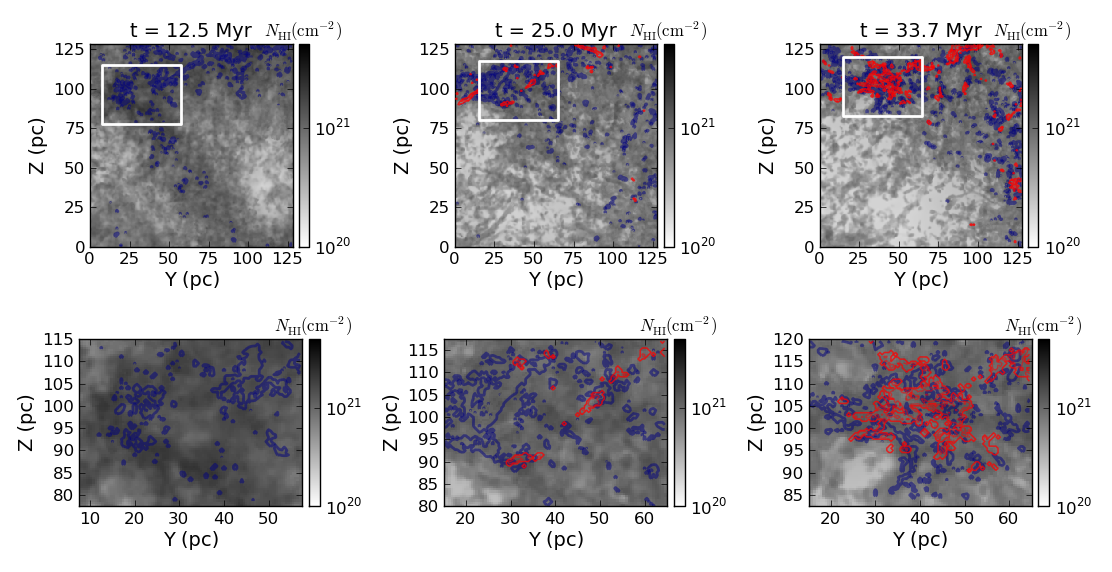}}
  \caption{\label{fig:hinsadepth} The spatial distribution of the atomic
  hydrogen column density (grayscale), the integrated CO intensity (thin
  red) and the depth of the dip caused by HISA (thick blue) are displayed
  for each timestep ({\it top panels}). 
  We used one HISA mask contour of 5 K and three
  $W_{\rm{CO}}$ contours of 0.1, 1 and 10 K km s$^{-1}$. The bottom
  panels show a zoomed-in section of the top panels. These sections are
  marked with white rectangles in the top panels. \textit{A color
  version of this figure is available in the online version of this
  journal.}}
\end{figure*}

After considering to what extent HISA strength maps can trace actual
HISA, we will now look at the connection between HISA and molecular
gas: Firstly, the CO gas. A one-to-one relation should not be expected, since HISA itself is
caused by atomic rather than molecular gas, but 
%perhaps 
it is generally thought that the presence of HISA is a sign of
atomic gas in the process of turning molecular. Therefore, HISA should
at least be expected to occur {\it near} molecular gas. 
%{\bf We compare the molecular gas column density $N_{\rm{mol}}$ to CO emission and our HISA measures, reminding the reader that $N_{\rm{mol}}$ is the column density calculated after determining whether the gas was molecular or atomic, but before determining  whether this gas is in the form of CO or not.}

In the top panels of Figure \ref{fig:hinsadepth} we show qualitatively how the
atomic gas ($N_{\rm{HI}}$, grayscale) and the
%molecular gas
CO emission ($W_{\rm{CO}}$, thin red contours) relate to the
HISA mask map (thick blue contours).  We remind the reader that no CO
emission was produced at the first timestep. In the bottom panels of
Figure \ref{fig:hinsadepth} we zoom into a smaller region (roughly the
same region 
% at each timestep
 in all three timesteps).
%to show in more detail how the HISA features are more or less bordering
%the molecular gas. 
It can be seen that the HISA mask dip depth contours
mostly border the CO emission contours as one might expect from
cold atomic gas moving towards sites where it turns molecular.

\begin{figure*}
  \resizebox{\columnwidth}{!}{\includegraphics{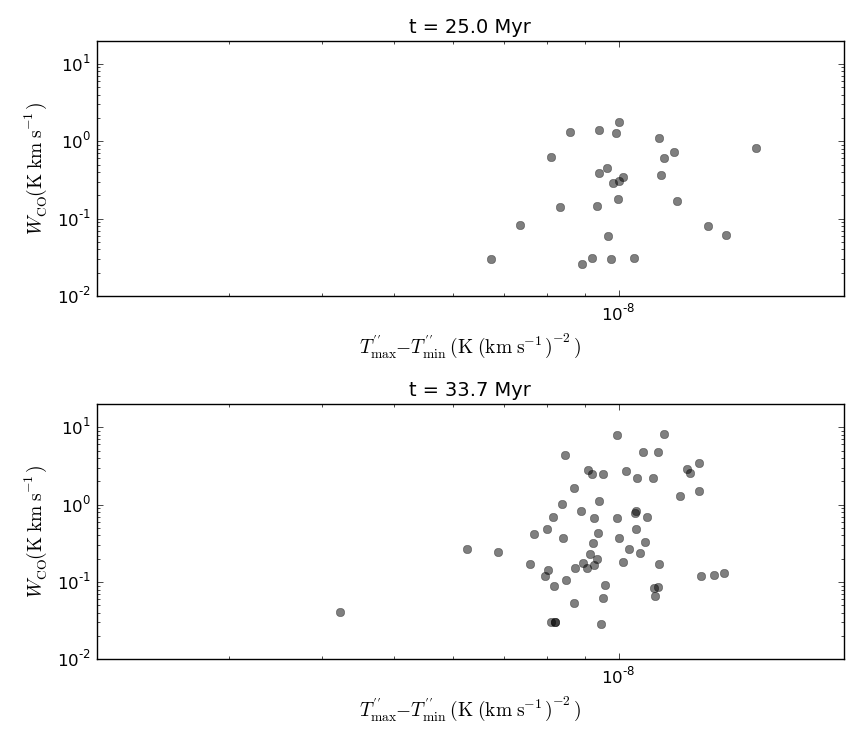}}
  \resizebox{\columnwidth}{!}{\includegraphics{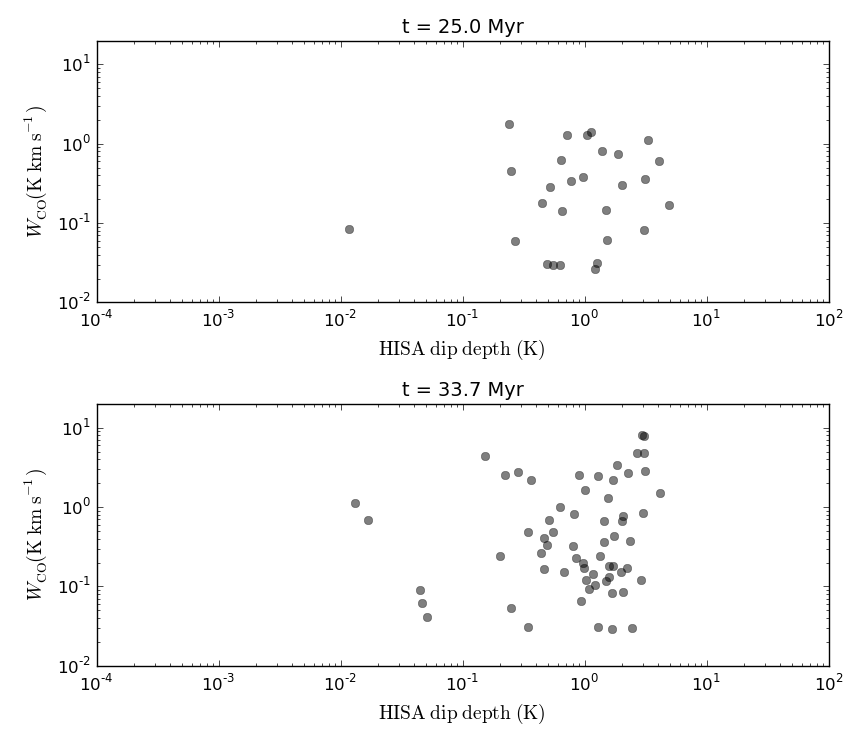}}
  \caption{\label{fig:strength_dip_WCO} Integrated CO emission
  $W_{\rm{CO}}$ against the HISA strength (original definition as used in \citet{2013MNRAS.429.3584H}, {\it left panel}) and
  HISA mask dip depth ({\it right panel}) using a pixel size of 8 pc$^2$. }
\end{figure*}

We compared the two HISA measures to the integrated CO emission
and found a very poor correlation (if any), even after averaging the data to $8 \times 8$-pc pixels, as shown in Figure
\ref{fig:strength_dip_WCO}. The averaging size was chosen based on the possible
correlation between the HISA features and the molecular gas: See the next Section.
Of course if the correlation is very poor as is the case here, smoothing of
any kind does not improve the correlation much.
There are relatively few points in these plots
%than in the ones with $N_{\rm{mol}}$
as only resolution elements above the
critical density of CO result in CO emission and therefore the CO gas
is a subset of the total molecular gas.

\subsection{Comparing HISA to the total molecular gas and CO-free gas}

\begin{figure*}
  \resizebox{\columnwidth}{!}{\includegraphics{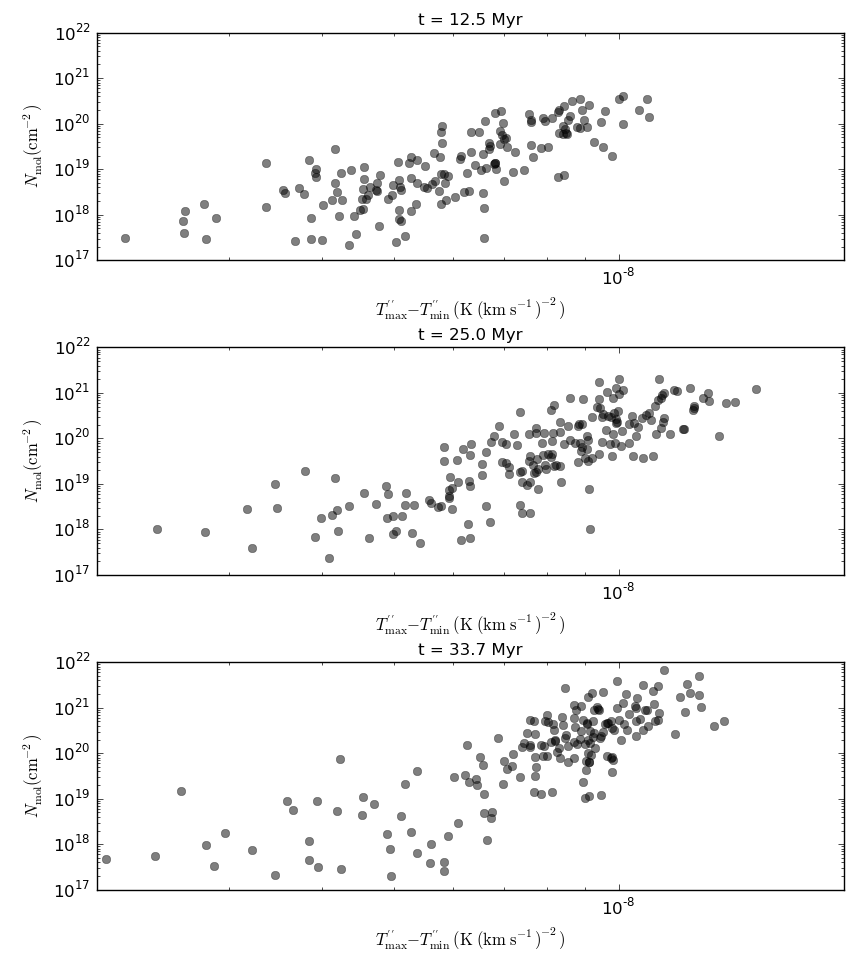}}
  \resizebox{\columnwidth}{!}{\includegraphics{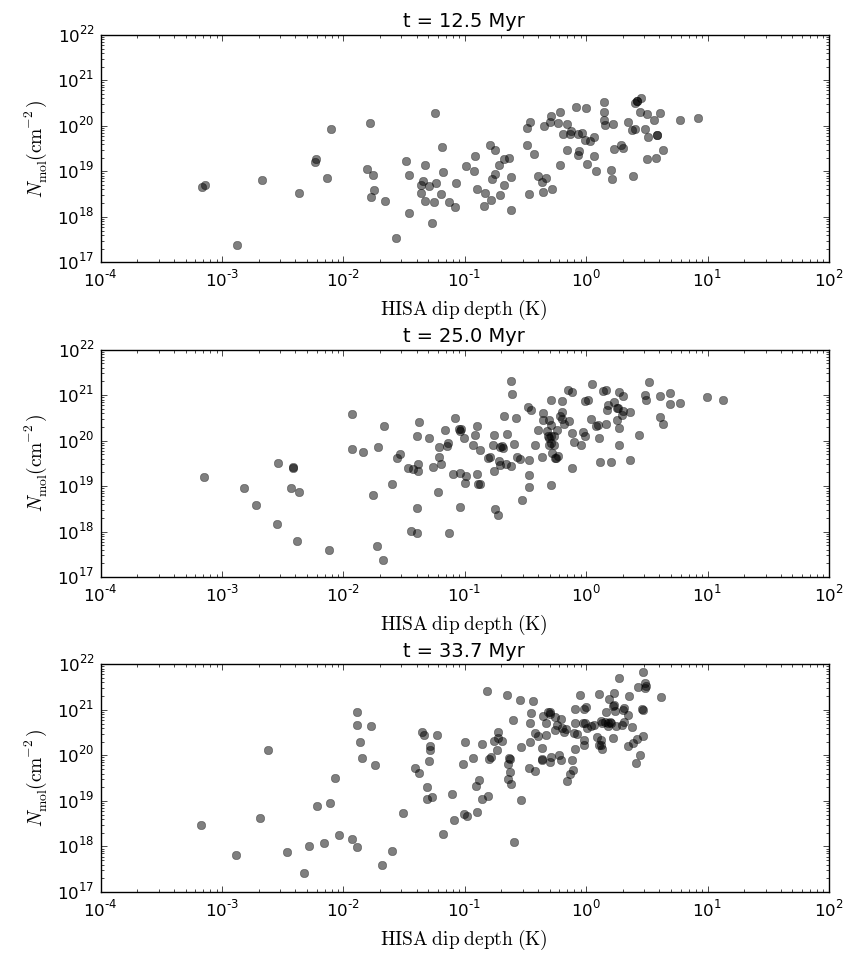}}
  \resizebox{\columnwidth}{!}{\includegraphics{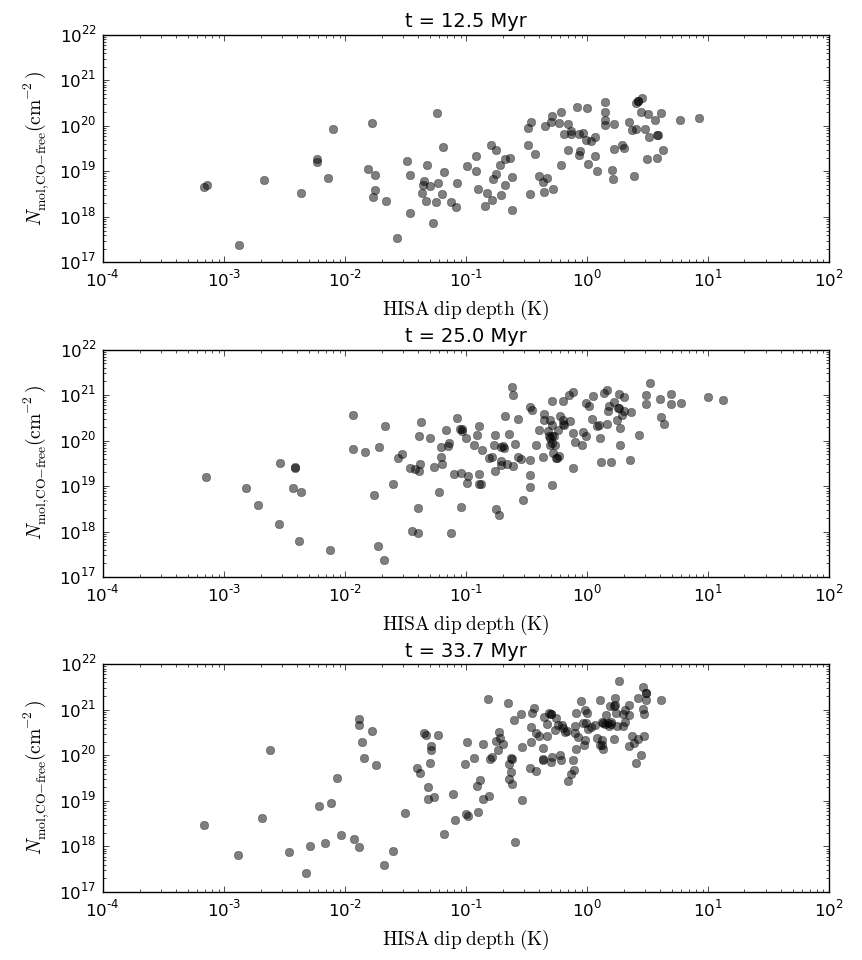}}
  \caption{\label{fig:strength_dip_Nmol} Similar to Figure \ref{fig:strength_dip_WCO}, we show the molecular gas column
  density against the HISA strength ({\it top left panel}) and the HISA
  dip depth ({\it top right panel}) using a pixel size of 8 pc$^2$. The
  CO-free molecular gas column density is shown against the HISA dip depth in the {\it lower panel}, using the same pixel size.}
\end{figure*}

\begin{table}
  \centering \begin{tabular}{llll} \hline\hline Time (Myr) & Res. (pc) &
  $r_{\rm{strength}}$ & $r_{\rm{match}}$ \\ \hline 12.5 & 1 & 0.38 &
  0.19 \\ & 2 & 0.44 & 0.26 \\ & 4 & 0.53 & 0.35 \\ & 8 & 0.65 & 0.55 \\
  & 16 & 0.75 & 0.79 \\ \hline 25.0 & 1 & 0.33 & 0.20 \\ & 2 & 0.40 &
  0.26 \\ & 4 & 0.48 & 0.33 \\ & 8 & 0.56 & 0.50 \\ & 16 & 0.64 & 0.59
  \\ \hline 33.7 & 1 & 0.26 & 0.17 \\ & 2 & 0.33 & 0.25 \\ & 4 & 0.42 &
  0.42 \\ & 8 & 0.50 & 0.61 \\ & 16 & 0.54 & 0.72 \\ \hline
  \end{tabular} \caption{\label{tab:pearsonr} Pearson $r$ correlation
  values as a function of increasing pixel size for the HISA strength
  ($r_{\rm{strength}}$) and HISA mask map ($r_{\rm{mask}}$).}
\end{table} 

%Having constructed HISA strength, HISA 
%match 
%mask and $N_{\rm{mol}}$ maps, we can compare how well they correlate, 
We now compare the HISA strength and HISA mask maps to the $N_{\rm{mol}}$ maps, keeping in mind that the correspondence between HISA strength and HISA
%match 
 mask is far from perfect (see the success rate plots, Figs.\
\ref{fig:successrate} and \ref{fig:successrate_blank}). 
In the top left panel of Figure \ref{fig:strength_dip_Nmol}, we show 
%that 
the HISA strength measure 
%correlates with 
against the molecular gas column density (the quantity derived directly
from the simulations as described previously), having averaged
the data to $8\times 8$-pc pixels. Averaging over fewer pixels worsens
the correlation and at the native 0.5 pixel separation there is barely
any correlation visible. This indicates that indeed there
is a correlation, although only at the level of neighboring positions,
rather than at the same projected position in the POS.

The top right panel of Figure \ref{fig:strength_dip_Nmol} shows the same
correlation, but using the HISA dip depth map values. It can be seen that
deeper dips tend to correspond to a higher molecular column
density. Taking these two figures together, it can be noted that both
the HISA strength and the HISA mask methods seem to show a similar weak
correlation, even though the HISA strength and HISA mask maps are not
identical.

Table \ref{tab:pearsonr} lists the correlation coefficients for various
pixel sizes. 
Pearson's {\it r} correlation coefficient is a measure of how strongly
two quantities are correlated, with a value of 0 meaning no linear correlation,
and a value of 1 meaning a perfect linear correlation. We used the {\it SciPy}
library\footnote{\url{http://www.scipy.org}} to compute this coefficient, which uses the standard definition
of Pearson's {\it r}.
%A value of 0 means no linear correlation, whereas a value
%of 1 signifies a perfect linear correlation. 
We computed
$r(\rm{strength},$ $N_{\rm{mol}})$ and $r(\rm{depth},$ $N_{\rm{mol}})$
for each timestep and pixel size, where the pixel values were calculated
as averages over the number of native pixels (0.5 pc)$^2$ contained
within each pixel. It can be seen that the correlation coefficients
increase with increasing pixel size.
%showing barely any correlation 
%increasing to a weak but visible correlation at the $\sim$ 10 pc cloud
%scale, where stronger HISA features roughly correspond to higher
%molecular gas column densities.

\begin{figure*}
  \resizebox{2\columnwidth}{!}{\includegraphics{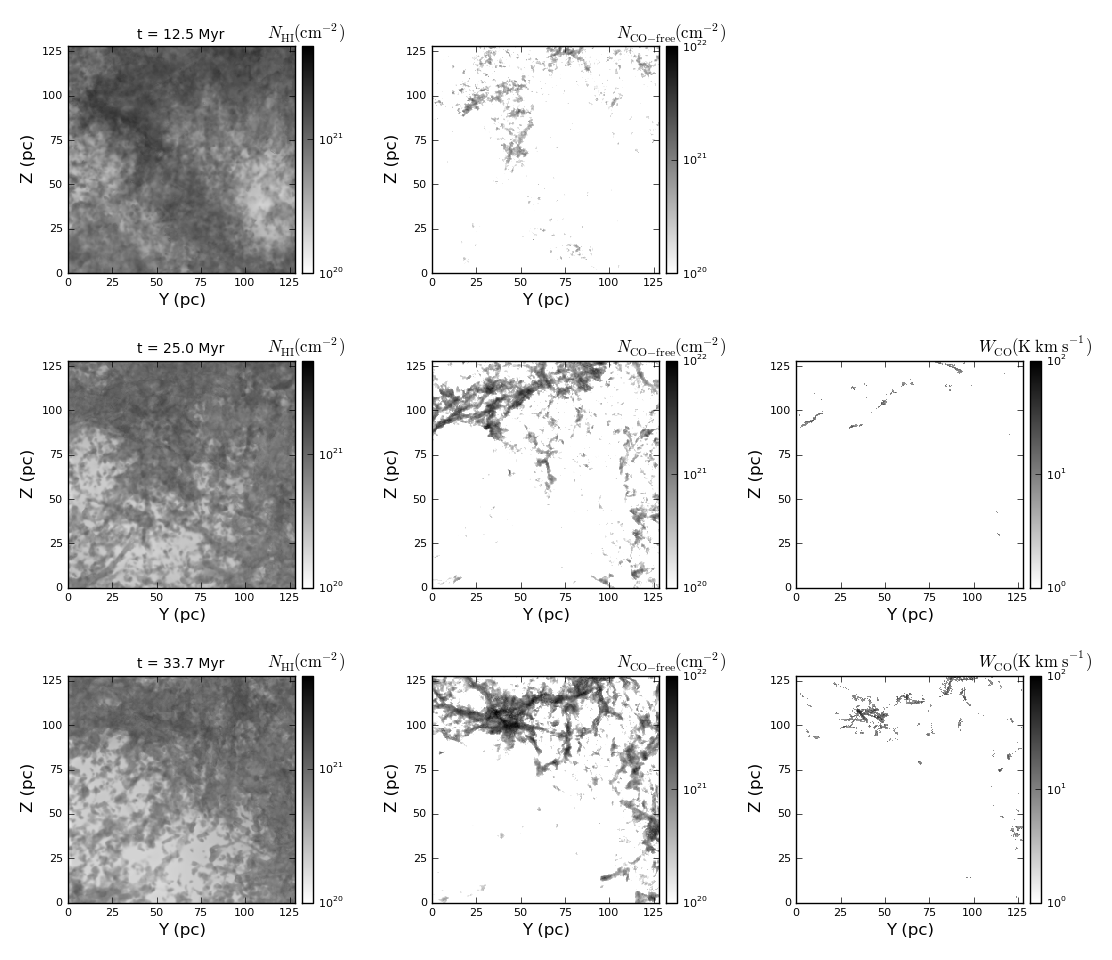}}
  \caption{\label{fig:NHI_NCOfree_WCO} {\it Left column:} Atomic hydrogen 
  column density. {\it Middle column:} CO-free molecular gas column density. {\it Right column:} Integrated CO intensity. The CO-free molecular gas
  column density is computed from the gas that is molecular, but not determined
  to be CO-emitting.}
\end{figure*}

For completeness, we show in the bottom panel of Figure \ref{fig:strength_dip_Nmol} the correlation between the CO-free gas column density, which is virtually indistinguishable from the plots with the total molecular gas column density. The reason for this is that the CO-emitting gas constitutes a small enough fraction of the total molecular gas content (in terms of particle number density) to be unnoticeable on a logarithmic scale. Qualitatively, we show the maps of the CO-free and CO-emitting gas in Figure \ref{fig:NHI_NCOfree_WCO} next to the atomic hydrogen column density map.  
%mass_sigv.py computes masses and velocity dispersions
%t=96: all gas 'CO-free' (9181 Msun)
%t=192: CO-free 46909; CO 3679
%t=259: CO-free 83719; CO 22126
The percentages of CO-free gas mass relative to the total molecular gas mass are dropping steadily at each successive timestep: Respectively, 100, 93 and 79 per cent. These values could in principle be compared to the dark gas fractions reported recently by \citet{2014arXiv1403.1589S}, although their simulations were set up differently and
also included explicit modeling of the gas chemistry.

\section{General structure of the clouds}

\subsection{Evolution of the PDFs}

\begin{figure*}
  \resizebox{0.9\columnwidth}{!}{\includegraphics{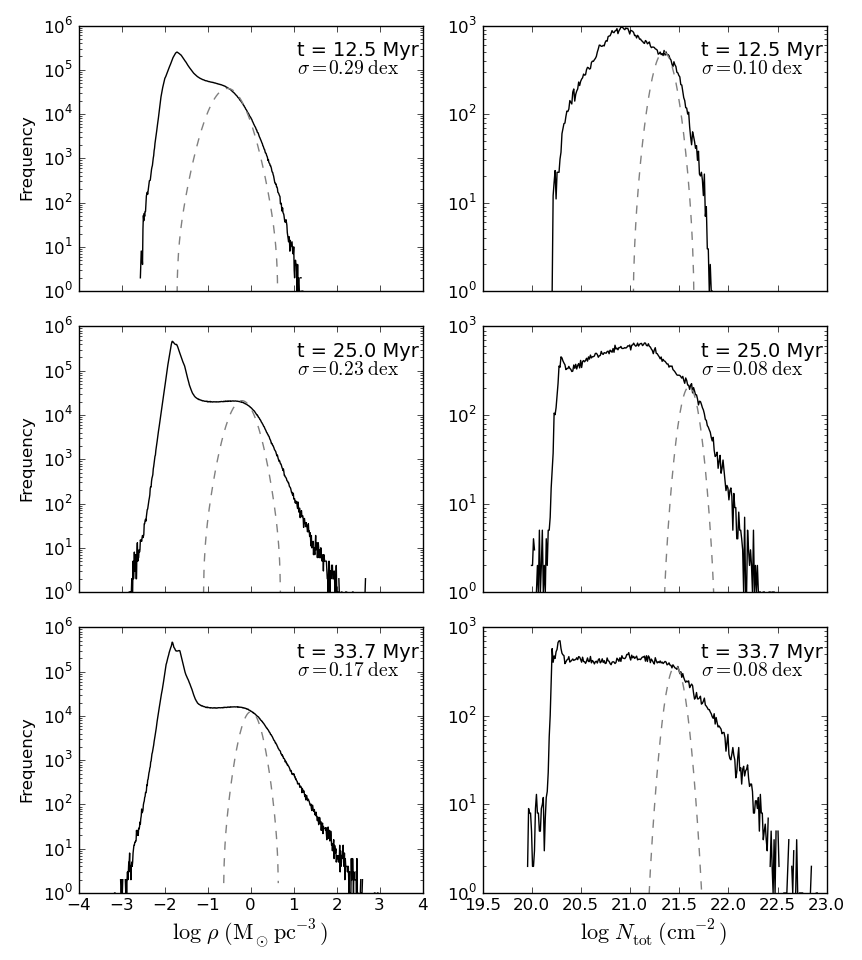}}
  \resizebox{0.9\columnwidth}{!}{\includegraphics{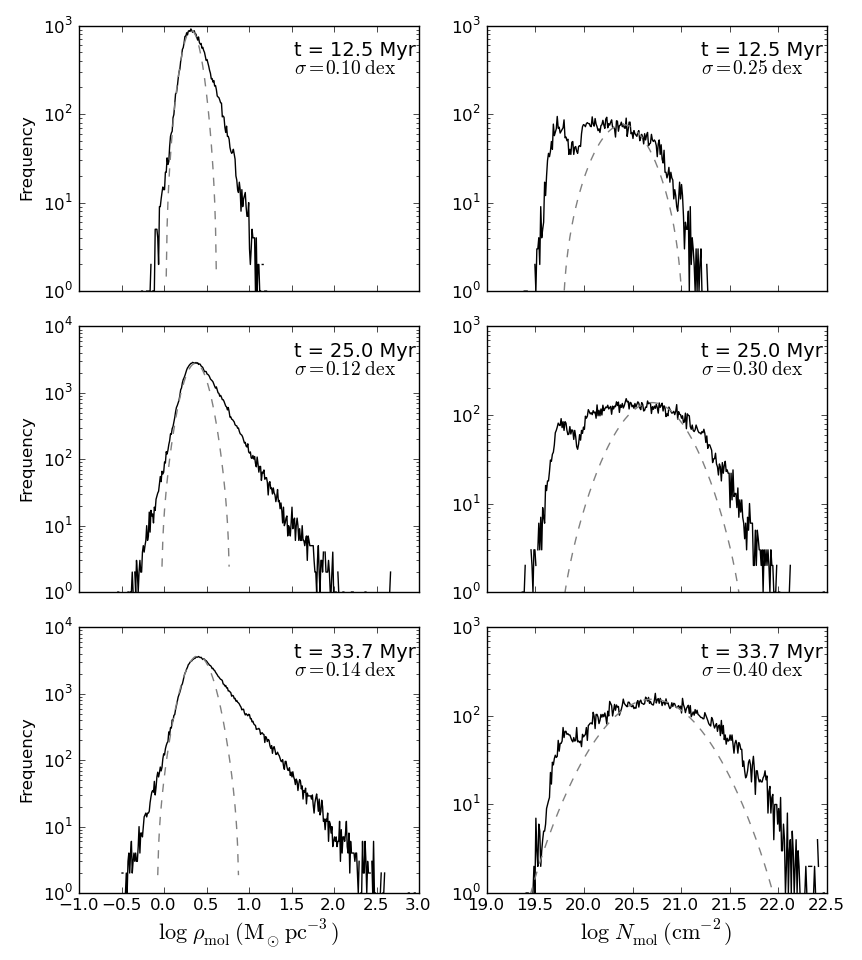}}
  \caption{\label{fig:PDFs} %For each timestep, we plot the 
  Volume- and column-density PDFs at all three timesteps
  studied in the simulation. The {\it left panels} shows the total gas PDFs,
  whereas the {\it right panels} shows the molecular gas PDFs. By-eye fits of
  the gaussian component of each PDF are shown with dashed lines draw
  attention to the qualitative structure of the PDFs. }
\end{figure*}

\begin{figure*}
  \resizebox{0.9\columnwidth}{!}{\includegraphics{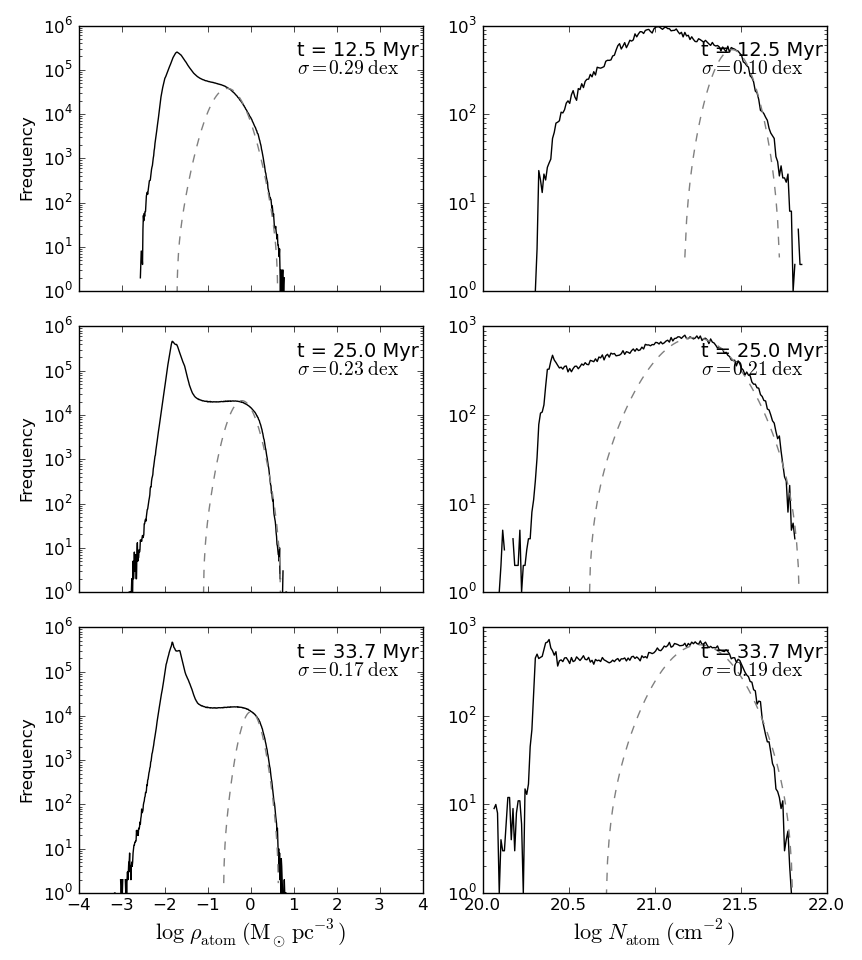}}
  \resizebox{0.45\columnwidth}{!}{\includegraphics{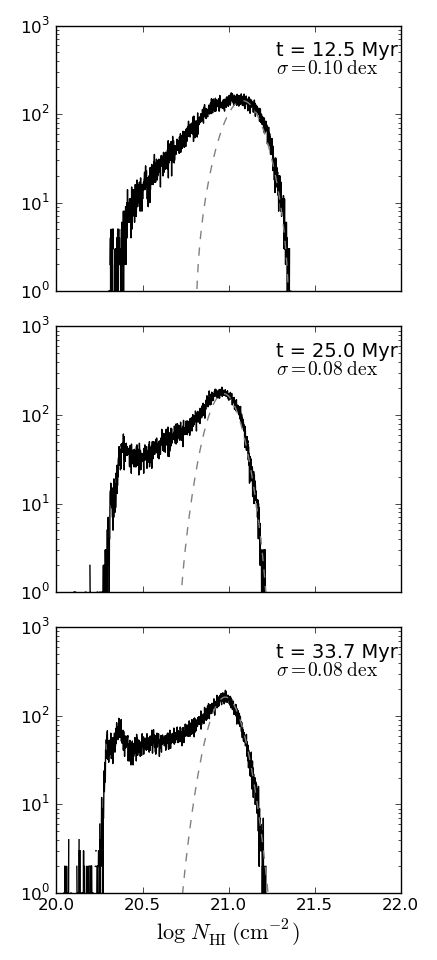}}
  \caption{\label{fig:PDFs_atom} Volume density PDFs of the
  atomic gas {\it (left column)}, and column density PDFs of the atomic
  gas, either derived directly from the simulation ($N_{\rm{atom}}$,
  {\it middle column}) or derived from the synthetic \HI\ line
  profiles ($(N_{\rm{HI}})$, {\it right column}). By-eye fits of the 
  %gaussian 
  log-normal component of
  each PDF are shown by the dashed lines.}
\end{figure*}

In order to show the distribution of the gas density in the simulation
volume, we produced the (volume weighted) volume ($\rho$)- and column ($N$) density 
probability density functions, as plotted in Figures \ref{fig:PDFs} and
\ref{fig:PDFs_atom}. In all cases we fit (by-eye) a log-normal gaussian
to the higher density end of the distribution to draw attention to the qualitative shape of the
PDFs.

%{\bf (JONATHAN: I would fit the lognormals to the {\it right} wing of
%the PDFs, not the left, as it is at the high densities that the gas
%behaves nearly isothermally, a prerequisite for a lognormal PDFs.)}

%In the following, we explain the general shape of the PDFs based on
%\citet{2011MNRAS.416.1436B}. 
We note that all $\rho$-PDFs, except those for the molecular
gas, show the characteristic bimodal shape of thermally bistable
gas \citep{2000ApJ...540..271V, 2005ApJ...630..911G,
2013ApJ...765...49G, 2005A&A...433....1A}.
%a low-density peak that is caused by the 
%general background density level. 
 A corresponding feature is also observed in the
$N$-PDFs, although with a much more lower amplitude.

%{\bf (JONATHAN: are your PDFs volume- or mass-weighted? Mass weighting
%could explain the reduction of the WNM peak in the PDFs.)}

The log-normal shape of the $\rho$PDFs is expected to occur for nearly
isothermal flows
when no star formation is happening and the 
%gas clouds are
flow is turbulence-dominated. In our case this is particularly obvious in the
first timestep, when molecular gas already exists but is not enough to
result in CO emission.  Conversely, in the two last timesteps, when
gravity has become the dominant driving force,
%Once star formation starts taking place 
a power-law tail 
%develops 
 is seen to have developed at the high density end of the PDFs
%. We see in Figure \ref{fig:PDFs}, left two columns
(Figure \ref{fig:PDFs}, first and third columns).
%, that we indeed see a
%progressively more dominant power-law tail. 
This is caused by the molecular gas part of the gas, as can be seen in
the third column of Figure \ref{fig:PDFs}, whereas the PDFs in
Figure \ref{fig:PDFs_atom} (depicting the atomic gas) show no obvious
power-law tails. It should also be noted that the power-law tail in the
molecular column density PDFs (Figure
\ref{fig:PDFs}, right two columns) is not 
%recovered 
so prominent due to line-of-sight
confusion.
% \citep[e.g.][]{1998PhRvE..58.4501P}.
 Although not a conclusive proof, these results are consistent with,
 and strongly support, the scenario that molecules themselves are not
necessary for gravitational collapse, and that instead they are {\it formed}
as a consequence of collapse in all three directions allowing
a visual extinction above unity to be achieved \citep{2001ApJ...562..852H,2008ApJ...689..290H,2012MNRAS.421....9G}.

Figure \ref{fig:PDFs_atom} also shows both the atomic gas N-PDF
(computed directly from the simulation; middle column) 
%as well as 
and the column density PDF ($N_{\rm{HI}}$-PDF) derived from
the synthetic observations (right column). It can be 
seen that the column densities derived from the synthetic observations
have a smaller dynamic range: The gas column becomes optically thick 
at a value of several times $10^{21}\ \rm{cm}^{-2}$, which in
turn causes \HI\ self-absorption features to appear in the line
profiles. Conversely, the atomic column density shows a larger
dynamic range since it is computed directly from the simulation before
the effects of self-absorption are taken into account. The column density
values are limited because the gas ultimately becomes molecular.
%{\bf (JONATHAN: this is not obvious at all from the plots. The
%two sets of histograms look qualitatively the same to me. Do you
%conclude this because the middle ones extend to higher values of $N$
%than the ones in the right column? In this case, I would not describe
%the situation as a ``cutoff'', but rather as a shorter dynamic range. A
%``cutoff'' suggests to me an abrupt dropoff, which is not apparent in
%the plots. Also, if this is what you mean, it would help showing the
%middle and right columns with the same range in the horizontal axis, in
%order to make the difference more evident.)}
Otherwise the profiles look qualitatively the same: a
low-density background-level peak and a log-normal shape.

%Finally, \citet{2001ApJ...557..727V} demonstrate a transition of the PDF
%from a log-normal shape to Gaussian form with an exponential
%intermediate stage, but a detailed study of the behaviour of the PDFs is
%beyond the scope of this work, and we did not follow the evolution of
%the clouds long enough.

\subsection{Accretion onto molecular clouds and kinematics}
%\subsection{Evolution of the gas and structure of molecular clouds}

\begin{figure*}
  \resizebox{2\columnwidth}{!}{\includegraphics{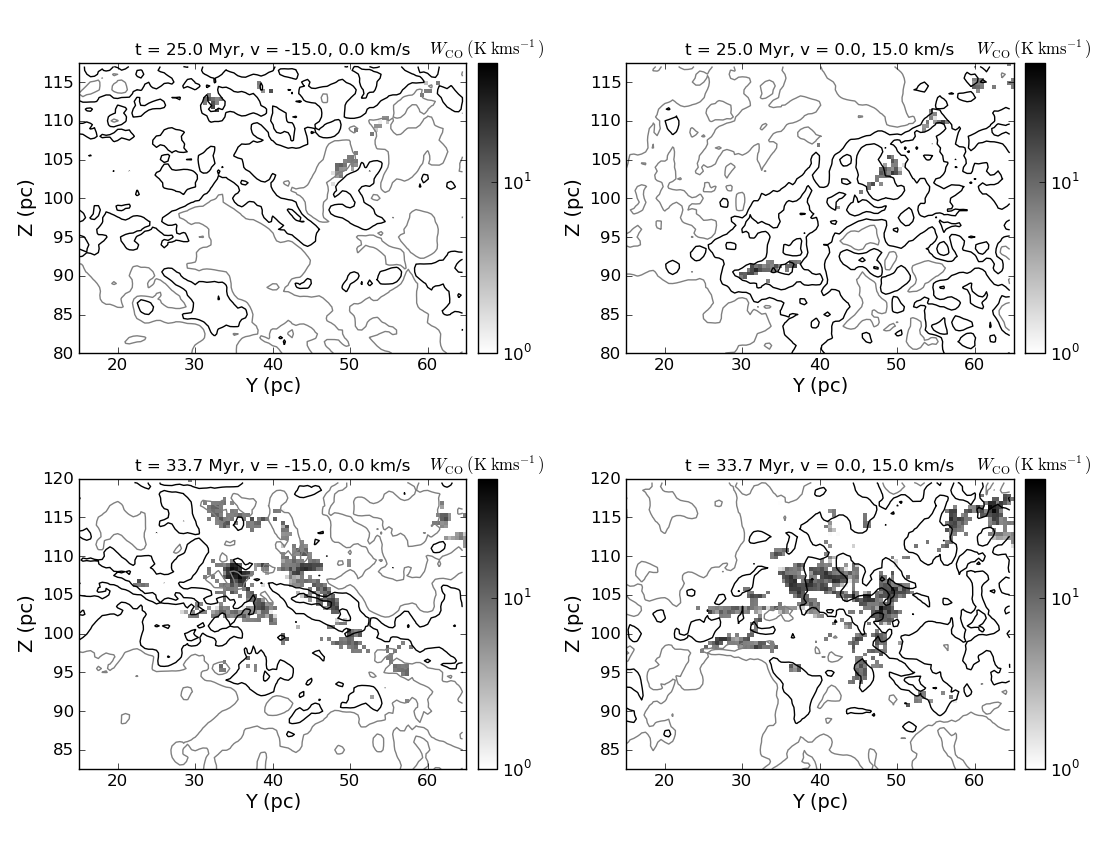}}
  \caption{\label{fig:Tb_ctr_TbCO_gray} Thick slices in velocity space
  (-15 to 0 and 0 to 15 km/s) of the \HI\ and CO brightness temperatures
  are shown in contours and grayscale respectively, integrated to
  $N_{\rm{HI}}$ and $W_{\rm{CO}}$ respectively. The \HI\ contours are 4,
  6, 8, 10, 12 and 14 $\times 10^{20}\ \rm{cm}^{-2}$, with the lowest
  contour level colored gray. Only the two later timesteps are shown
  since no CO emission was present in the first timestep.}
\end{figure*}

As discussed in Sec.\ \ref{sec:gral_morph}, the dense molecular
clouds in the simulation grow by accretion of \HI\ gas. It is thus
important to investigate the signatures of this process in the
observational domain.
Figure \ref{fig:Tb_ctr_TbCO_gray} shows the \HI\ and CO brightness
temperature (contours and grayscale
respectively) at the last two
% of the three 
timesteps (top and bottom rows, respectively), 
%and in three velocity slices of -3, 0 and 3 $\rm{km\s^{-1}}$. 
and in two velocity intervals, from $-15$ to $0\, \kms$ and from 0 to
$+15\, \kms$ (left and right columns, respectively).
%For each timestep, a different region was chosen to center on a
%particular cloud structure. 
Different subregions of the simulation were chosen in each timestep, so
that each was centered on a prominent cloud structure at the given
timestep.  
%At the 
The first timestep ($t=12.5$ Myr), is not shown because, as
mentioned above, no CO emission 
%can be seen since the gas did not exceed the critical density, whereas
%more CO emission developed quickly at the later timesteps. 
is produced at that time.

%Due to velocity crowding along the line of sight,
%confusion (the gas velocity does not correspond one-to-one
%to the spatial line-of-sight distribution due to the motions of the gas)
There is only a very vague structure visible of CO gas surrounded by
atomic gas, possibly due to projection effects. We also note that channel maps such as these may show more
small-scale structure than the three-dimensional maps, or slices through these maps. The reason for this is that there is no strictly uniform relation between the velocity and
the spatial axis, a certain amount of randomization of the structure
takes place \citep{2000ApJ...532..353P}.

%Pichardo et al. 2000 also say that the channel maps should resemble the spatial distribution of the LOS velocity (but weighted by the density) rather than the density field.

%\textit{Here we should show overlay plots of a particular filament/cloud followed in every timestep; we should have one.}

\begin{figure*}
  \resizebox{0.7\columnwidth}{!}{\includegraphics{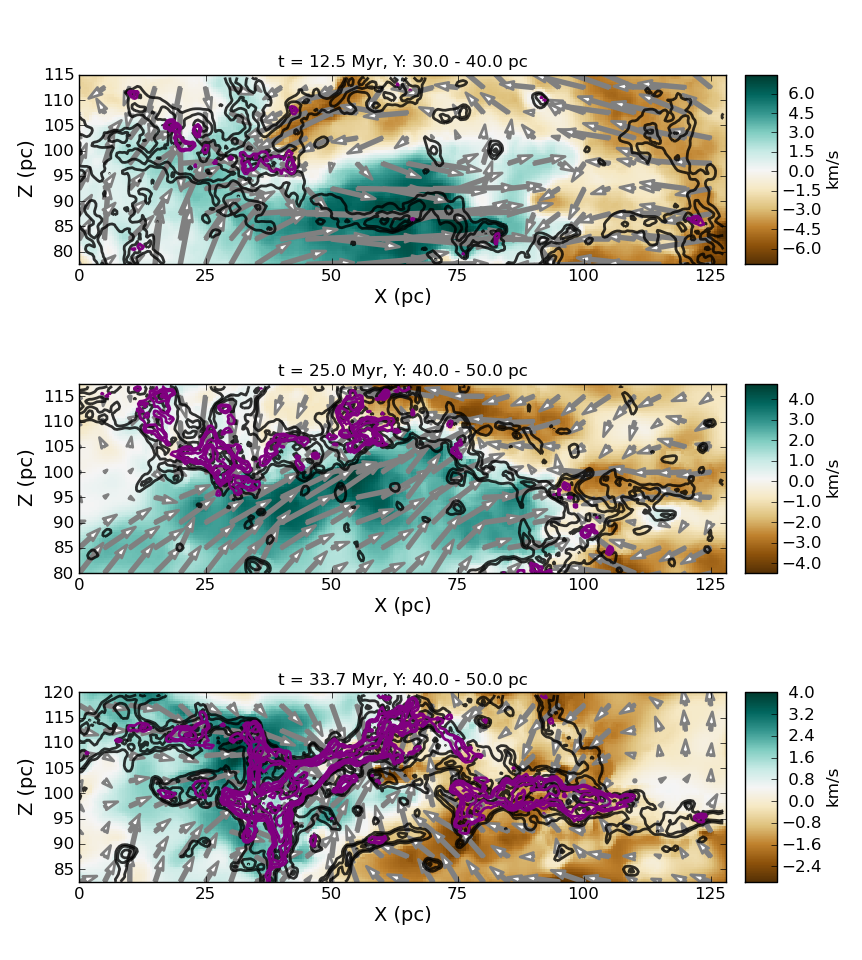}}
  \resizebox{0.6\columnwidth}{!}{\includegraphics{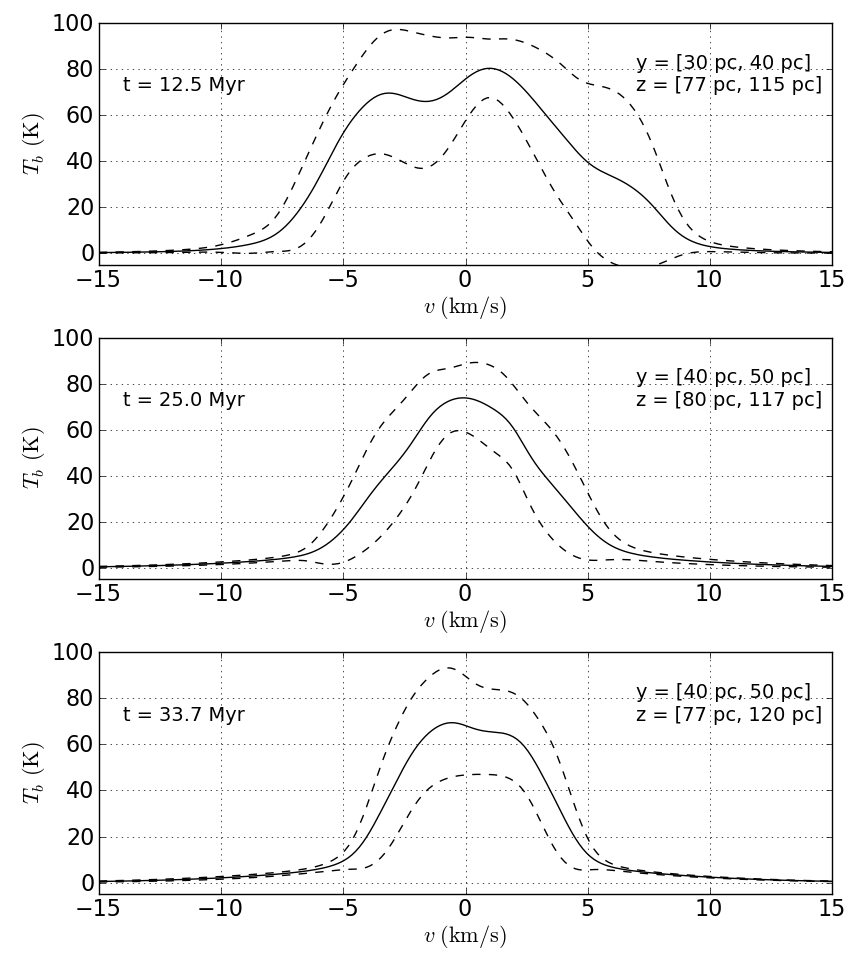}}
  \resizebox{0.6\columnwidth}{!}{\includegraphics{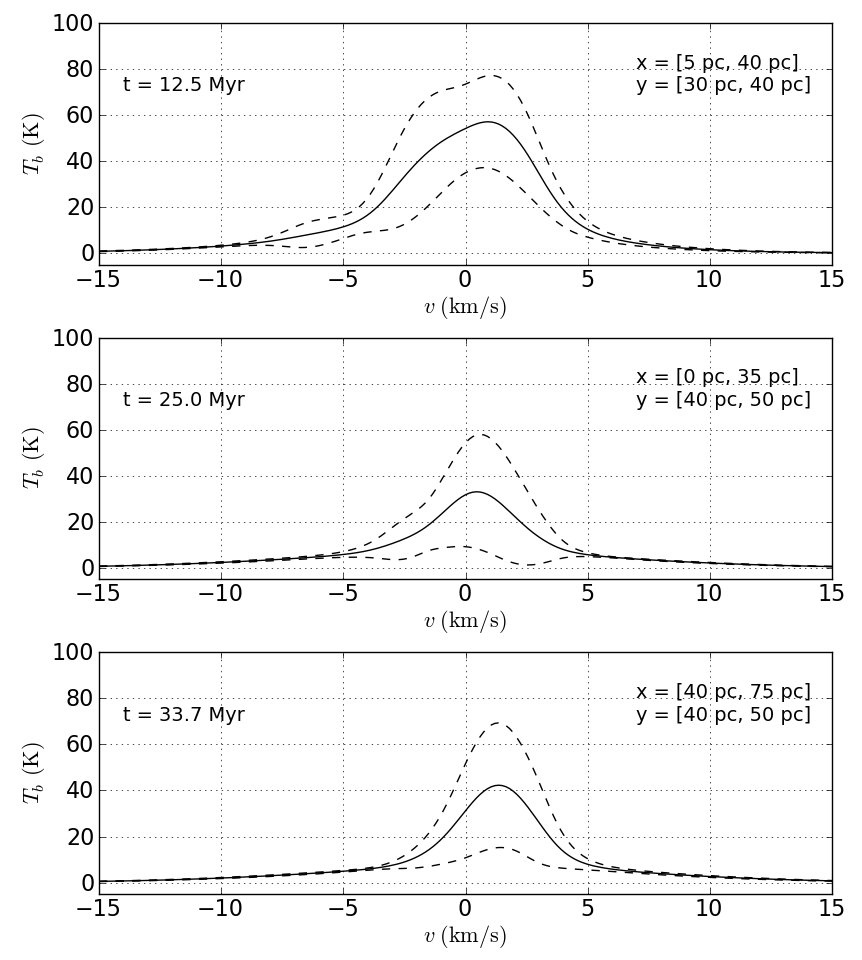}}
  \caption{\label{fig:xvx} %Here we show slices in the x-direction 
  {\it Left panels:} Slices of the atomic (thin black contours) and molecular (thick
  purple contours) gas density on the $(x,z)$-plane in a subregion of
  the simulation containing an evolving massive cloud, shown at the
  three timesteps investigated throughout the paper. The arrows show the
  projection of the velocity vector on this plane, and the color scale
  depicts the value of the $x$-component of the velocity on the plane. A
  positive value of this component means a flow from left to right in
  the figure. \textit{A color version of this
  figure is available in the online version of this journal.} 
% to illustrate the gas flow (left panel). The
% velocity in the x-direction is shown in colorscale, where a positive
% velocity means a flow from left to right and a negative velocity means
% a flow in the opposite direction. The arrows indicate the combined x,
% z direction of the velocity field. Black thin contours indicate the
% atomic gas column density (integrated over the y direction range),
% whereas thick purple contours mark the molecular gas column
% density. 
  In the {\it middle and right panels}, we show the \HI\ brightness
  temperature profiles resulting from these regions, as observed
  from the positive-$x$ direction (middle panels) and from the positive
  $z$ direction (right panels). It is seen that the peaks of the \HI\
  emission correspond to the typical values of the \HI\ velocity
  approaching the molecular gas. As before, the dashed lines mark
  1-$\sigma$ deviations from the average line profile. Note that
  the profiles in the $z$-direction are less pronounced because they
  span the full $z$-coordinate range, but the selected gas structures
  as shown in the left panels only span a small fraction of the 
  z-coordinate range.}
\end{figure*}

%In order to demonstrate that the double peaks in the \HI\ profiles are
%caused by colliding gas flows, at least in our simulation, 

In Figure \ref{fig:xvx} we plot cuts of the density field
%in the x-direction. 
of both the \HI\ (thin black contours) and the molecular (thick purple
contours) gas density on the $(x,z)$-plane, as well as the projection of
the velocity vector on this plane (grey arrows) and the $x$-component of
the velocity field (color scale). 
%In the figures for each
%timestep, the colorscale shows the velocity component in the x-direction
%and the arrows show the velocity field in the x, z direction. 
The green colors indicate a positive velocity, from left to right along
the x-axis, whereas the brown colors indicate the opposite direction. In
the middle and right panels of Figure \ref{fig:xvx} we show the line
profiles extracted from the same regions, as seen by an observer at
the right of the fields (middle panels) or from above them (right
panels). It is clear from these figures that the \HI\ gas is in the
process of
%colliding
accreting onto the dense molecular cloud.

The top-left panel of Figure \ref{fig:xvx} (at $t=12.5$ Myr) shows
that the dense structure on the left side of the field, centered at $(x,z)
\approx (25, 100)$ pc, and which contains both
\HI\ and (CO-free) molecular gas, is accreting diffuse \HI\ material
predominantly along the vertical ($z$) direction. There is also a
converging flow along the $x$-direction, with the collision region
centered roughly at $(x,z) \approx (90, 90)$ pc, but this flow seems to
not have formed a dense cloud yet, and instead it seems to be {\it
carrying} previously formed, predominantly atomic clouds spanning the
ranges $(\Delta x,\Delta z) \approx (50$:$80,80$:$90)$ pc and $(\Delta
x,\Delta z) \approx (110$:$130,80$:$110)$ pc, respectively. In the
middle-left panel ($t=25$ Myr), the flow seems to have joined the two
components described above into a single, long filament stretching along
most of the $x$-extension of the region, with the positive (blue) and
negative (brown) components of the $x$ velocity being located below-left
and above-right of the cloud complex, respectively. The complex is
highly molecular at the left, but still mostly atomic at the right part
of the field. Finally, in the bottom-left panel, the complex is seen to
have become mostly molecular, and is still accreting diffuse \HI\ material
mostly along the vertical ($z$)-direction. Although not shown here, at
this time the cloud is copiously forming stars.

In the middle and right panels, we note that the accretion process often
results in double-peaked \HI\ line profiles. Such features are common in
observations of \HI\ in the neighborhood of molecular clouds (see, e.g.,
the top panel of Fig.\ \ref{fig:compare_profile}), and our results from
this Section suggest that such bimodal profiles are often the result
of the \HI\ flow onto the molecular clouds.

%Even though the final timestep shows molecular gas present in the
%colliding gas flows, at the two earlier timesteps this is not evident
%and in fact it appears that the molecular gas is present where the gas
%collides. This represents a different stage in the evolution of the
%molecular cloud. 
%
%While we interpret our results in the context of
%colliding gas flows (forming molecular clouds in the process), our
%simulation allows in principle the destruction of the molecular
%clouds. If gas in a particular cloud is not gravitationally bound, it
%can leave the cloud and reach an area where the visual extinction drops
%below unity. At that point we would consider the gas as once again
%atomic. Physically, this would be an effect of photodissociaton of the
%molecular gas once it leaves the shielded areas of the molecular
%cloud. While we do not see any clear evidence of this happening in our
%simulation, we did not track the gas particles in order to see how often
%this scenario happened.

\section{Discussion and conclusions} \label{sec:concl}

%In the future, we intend to investigate the detailed properties of the
%clouds present in this simulation, tracking their evolution through
%time.  Additionally, the physical properties of the ``void'' in our
%simulation volume can be studied. %Theory \textit{(ref? -- at least in
%%the cosmological context)} predicts that voids become more spherical as
%%time passes and gas evacuates the void.

\subsection{Limitations} \label{sec:limitations}

One of the key missing features in the present study 
%that should have an important effect on the cloud structure, 
is the inclusion of supernova (SN) feedback, which should maintain
the turbulence driving at the scales of our simulation. This may, in
turn, have a significant effect on the evolution and structure of the
clouds, although its relative importance compared to the gravitational
driving of the motions in the clouds is uncertain. Supernovae tend
to explode in regions that have been previously evacuated by ionizing
radiation and winds from the massive stars, and numerical simulations of
this scenario (albeit without self-gravity) suggest that the dense
clouds are not strongly affected by the supernovae \citep[e.g.,][]
{2004A&A...425..899D, 2012ApJ...750..104H}. In a future study, we plan to
repeat our analysis in the presence of ionization and SN feedback, as
well as the magnetic field and ambipolar diffusion.
%in order to simulate \HI\ (super-)shells, which are
%commonly seen in galactic disks
%\citep[e.g.][]{1979ApJ...229..533H}

%One of the most 
Another important and obvious improvement we can make to our model
is a more detailed treatment of the formation and destruction of
molecular gas \citep[cf. e.g.][]{2014arXiv1403.1589S}. 
For example, we assume that the formation of
molecular gas happens on a timescale that is very small relative to the
timesteps we considered, but if we follow the evolution of the gas
particles more closely a minimum timescale for molecular gas formation
can be imposed as well as a certain balance of photodissociation. 
As a consequence, the molecular and CO fractions in our simulation must
be considered as upper limits. At least our approach provides a 
first order approximation to the evolution of atomic and molecular gas.

\subsection{Summary} \label{sec:summary}

%Here, we 
We have presented a numerical simulation where `molecular' clouds 
%form 
are identified by using a
\textit{post-facto} processing of the particles in the simulation
volume: molecular gas was assumed to have formed if the local
temperature had dropped below 50 K and the local average extinction,
%had risen above 
$\Av> 1$. We also created synthetic observations of \HI\ and CO,
the latter assumed to exist at a grid cell if the conditions for
molecular gas were satisfied and besides the local density satisfied $n
> n_{\rm crit}$, where $n_{\rm crit}$ is the critical density for CO
formation (cf.\ Sec.\ \ref{sec:synth_obs}).
% where we note that (particularly in the first
%timestep under consideration) not all the molecular gas will emit in CO
%due to the gas remaining below the critical density for CO emission.

%Then, 
We used two different methods to identify (potential) HISA
features. 
%First, 
One, we used the amplitude of the second derivative of the
\HI\ brightness temperature profile, dubbed `HISA strength'. 
%Then, 
Two, we
identified HISA features by matching local minima in the \HI\ brightness
temperature profile to peaks in the calculated opacity at the same
velocity (that we call `HISA 
%match
mask'), thereby distinguishing between dips in the profile caused
by HISA and those caused by the absence of atomic gas. Although this
method is superior because it uses local gas opacity information, it
cannot be applied to actual observations since this kind of information
is generally not available. Nevertheless, it provides a means of
testing the goodness of the first method. We then compared the
location and intensity of the HISA features with the presence of
molecular gas and looked at the structure of the gas. Finally, we
investigated the density and column-density PDFs of the gas, the latter
obtained both directly from the simulation data and from the synthetic
observations. We focused on three timesteps of the simulation, namely at
$t=12.5$, 25 and 33.7 Myr. The first timestep corresponds to a time when
the clouds are still developing and no SF is occurring in the
simulation. The second corresponds to a time when SF is at an early
phase, while the latter corresponds to a time when SF is copious. At
this time, the effects of stellar feedback are clearly missing and
should be included in future studies. 
%We summarize our conclusions
Our main results were as follows:

\begin{itemize}

  \item At the first of the timesteps studied,
  significant quantities of molecular gas exist, but no significant CO has
  formed yet. That is, the molecular gas is `CO-free'. Instead, at the
  last of the three timesteps, CO is abundant. This result is
  consistent with the scenario that gravitational contraction drives the
  formation of CO molecules \citep[cf.][]{2008ApJ...689..290H}.

  \item At the 
%native 
  chosen resolution of our 
%simulation 
  gridded data (0.5 pc pixel separation), there is a very poor
  spatial correlation between the HISA strength and the HISA 
%match,
  mask.
%which is only weakly related to thresholding of the HISA strength
%  (excluding lower values). 
  The correlation improves 
% after including HISA match pixels within increasing pixel radii.
  if matches with neighbouring pixels (in the projected plane of
the sky, POS) are allowed.

  \item However, 
%even though HISA strength and HISA {\bf mask} measures do
%  not agree very well at the native resolution, 
  both HISA indicators show a weak but significant correlation
  with the molecular gas column density on a scale of a few to several
  parsecs on the POS. At smaller scales no correlation is
visible. This suggests that HISA is located on the periphery of the
molecular emission, rather than coincident with it.

  \item The volume and column density PDFs extracted from the simulation
  show the expected transition from a purely log-normal shape to one
  with a power-law tail when molecular clouds form. However, the
power-law tail is only seen in the molecular components, and not in the
\HI, suggesting again that molecule formation is directly correlated with
gravitational infall.

  \item At least in our simulation, most of the multi-peaked \HI\ line
  profiles in the neighborhood of molecular clouds are caused by
  bulk \HI\ flows into the molecular clouds, rather than by \HI\
  self-absorption.
\end{itemize}

\section*{Acknowledgments}
JSH and the computer cluster on which the numerical simulations were
carried out are supported by CONACYT grant 102488 to Enrique
V\'azquez-Semadeni. 

\label{lastpage}
\bibliographystyle{mn2elong} %to fix some errors about entry string being too long
  %specifically Draine et al. 2007
%\bibliographystyle{mn2e}

\end{document}